

\input harvmac
\def\bs{\bigskip}

\def\no{\noindent}
\def\hb{\hfill\break}
\def\qq{\qquad}

\def\br{\bigr}

\def\IR{\relax{\rm I\kern-.18em R}}

\def\z1 {z^{-1}}
\def\z2 {z^{-2}}
\def\z3 {z^{-3}}
\def\z4 {z^{-4}}

\def \bab  {{\bar b }}
\def \sms {$\s$-models\ }
\def \tA  { {\tilde {A }}}
\def \tbA  { {\tilde {\A }}}
\def \tJ  { {\tilde {J }}}
\def \tbJ  { {\tilde {\J }}}
\def \be {\bar \ep }

\def \dim  {{\rm dim  }}
\def \bd {\bar \del}

\def \tt {{\tilde \t}}
\def \G {\Gamma} \def \k1 {{1\over
k}} \def \bh { {\bar h} } \def \ov { \over }

\def \O {\Omega }

\def \ra {\rightarrow}

\def \W { {\bar \o }}
\def \half{{\textstyle{1\over 2}}}

\def \D {\Delta}

\def \a {\alpha}
\def \b {\beta}

\def \Tr {{\ \rm Tr \ }}

\def \ln {{\rm \ ln \  }}
\def \det {{\ \rm det \ }}

\def \l {\lambda}
\def \1p {{1\over  \pi }}
\def \2p {{{1\over  2\pi }}}
\def \4p {{ {1\over 4 \pi }}}
\def \8p {{{1\over 8 \pi }}}
\def \P^* { P^{\dag } }
\def \p {\phi}

\def \m {\mu }
\def \n {\nu}
\def \ep {\epsilon}
\def\g {\gamma}
\def \r {\rho}
\def \k {\kappa }
\def \d {\delta}
\def \o {\omega}
\def \s {\sigma}
\def \t {\theta}

\def\mn {{\m\n}}

\def \Bmn {B_{\m\n}}
\def \fourth {{\textstyle{1\over 4}}}
\def \third {{\textstyle{1\over 3}}}

\def \e#1 {{{\rm e}^{#1}}}
\def \const {{\rm const }}

\def \eq#1 {\eqno {(#1)}}
\def \sm {$\s$-model\ }

\def \bd  {{ \bar \del }}

\def \D  {\Delta }

\def \V { {\tilde V } }
\def \E {{ \tilde E}}
\def \bH {{\bar H}}

\def \bd  { \bar \del }
\def \D  {\Delta }

\def \ov {\over }
\def \gG {{\rm g_G}}
\def \gH {{\rm g_H}}
\def \kg { k + \gG }
\def \kh { k + \gH }
\def \A  { {\bar A} }

\def \O  {{\Omega}}

\def \o {\omega}

\def \p {\phi}
\def \ep {\epsilon}
\def \s {\sigma}

\def \r {\rho}
\def \d {\delta}
\def \l {\lambda}
\def \m {\mu}
\def \g {\gamma}
\def \n {\nu}

\def \fourth {{1\over 4}}
\def \third {{1\over 3}}
\def \e#1 {{{\rm e}^{#1}}}
\def \const {{\rm const }}

\def \vp {\varphi}

\def \half { { 1\over 2 }}

\def \J {\bar J }
\def \P {\Phi}
\def\np {  Nucl. Phys. }
\def \pl { Phys. Lett. }
\def \mpl { Mod. Phys. Lett. }
\def \prl { Phys. Rev. Lett. }
\def \pr  { Phys. Rev. }

\def \cmp { Commun. Math. Phys. }
\def \ijmp { Int. J. Mod. Phys. }
\baselineskip6pt
\Title{\vbox
{\baselineskip 6pt{\hbox{CERN-TH.6969/93}}{\hbox{THU-93/25}}{\hbox
{Imperial/TP/92-93/59}}{\hbox{hep-th/9310159}} } }
{\vbox{\centerline { Antisymmetric tensor
coupling }\vskip2pt
 \centerline {and conformal invariance  in sigma models }\vskip2pt
 \centerline{
corresponding to gauged WZNW theories }
}}
\vskip -38 true pt
\centerline  { {  Konstadinos Sfetsos\footnote {$^\dagger$} {e-mail address :
sfetsos@ruunts.fys.ruu.nl} }}
 \smallskip
\centerline {\it  Institute for Theoretical Physics, Utrecht University }
\centerline {\it Princetonplein 5, TA 3508, The Netherlands}
\medskip
\centerline {and}
\medskip
\centerline{   A.A. Tseytlin\footnote{$^{*}$}{\baselineskip8pt
On leave  from Lebedev Physics
Institute, Moscow, Russia.
e-mail: tseytlin@surya3.cern.ch \ \  and tseytlin@ic.ac.uk} }
\smallskip
\centerline {\it Theory Division, CERN, CH-1211 Geneva 23, Switzerland}
\centerline  {and}
\smallskip
\centerline {\it Blackett Laboratory, Imperial College,  London SW7 2BZ, U.K.}
\medskip
\centerline {\bf Abstract}
\smallskip
\baselineskip3pt
\noindent
String backgrounds associated with  gauged $G/H$ WZNW models generically
depend on $\alpha'$ or $1/k$.
The exact expressions for the corresponding metric $G_{\m\n}$,
antisymmetric tensor $B_{\m\n}$, and dilaton $\phi$ can be obtained by
eliminating the $2d$ gauge field from
the local part of the effective action  of the gauged WZNW model.
We show that there exists a  manifestly gauge-invariant prescription for the
derivation of
the antisymmetric tensor  coupling and
discuss some subtleties involved. When the subgroup $H$ is one-dimensional and
$G$ is simple
the antisymmetric  tensor is given by the semiclassical ($\a'$-independent)
expression.
We consider in detail the  simplest non-trivial example
with $B_{\m\n}\not=0$ -- the $D=3$
$\sigma$-model corresponding to the $[SL(2,\IR)\times \IR]/\IR$ gauged WZNW
theory
(`charged black string') and show that
 the exact expressions for $G_{\m\n}$, $B_{\m\n}$ and $\phi$
solve the Weyl invariance conditions in the two-loop approximation.
Similar conclusion is reached for the closely related  $SL(2,\IR)/\IR$
chiral gauged WZNW model.  We find that there exists   a scheme in which the
semiclassical
background is also a solution of the two-loop conformal invariance equations
(but the tachyon
equation  takes a non-canonical form). We discuss in detail the role of field
redefinitions (scheme
dependence) in establishing a correspondence between the  sigma model and
conformal field theory
results.
\smallskip
\noindent
{CERN-TH.6969/93}

\noindent
 {October 1993}
\Date { }
\noblackbox
\baselineskip 20pt plus 2pt minus 2pt
\lref \bep {C. Becchi and O. Piguet, \np B315(1989)153. }
\lref \nov {S.P. Novikov,  Sov. Math. Dokl. 37(1982)3. }
\lref \fuch { J. Fuchs, \np B286(1987)455 and  B318(1989)631. }
\lref \sus { R. Rohm,  \pr D32(1985)2845.}
\lref \suss {   H.W. Braden, \pr D33(1986)2411.}
\lref   \red { A.N. Redlich and H.J. Schnitzer, \pl B167(1986)315 and
B193(1987)536(E);  A. Ceresole,
 A. Lerda, P. Pizzochecco
 and
P. van Nieuwenhuizen, \pl
 B189(1987)34.}
 \lref \div  { P. Di Vecchia,  V. Knizhnik, J. Peterson and P. Rossi, \np
B253(1985)701.}

\lref \schn {  H. Schnitzer, \np B324(1989)412.  }

\lref \all { R.W. Allen, I. Jack and D.R.T. Jones, Z. Phys. C41(1988)323. }

\lref \nem {D. Nemeschansky and S. Yankielowicz, \prl 54(1985)620; 54(1985)1736
(E).}

\lref \bep { C. Becchi and O. Piguet, \np B315(1989)153. }

\lref \ks {Y. Kazama and H. Suzuki, \np B321(1989)232; \pl B216(1989)112.}

\lref \mor {A.Yu. Morozov, A.M. Perelomov, A.A. Rosly, M.A. Shifman and A.V.
Turbiner, \ijmp
A5(1990)803.}

 \lref \tur {A.V. Turbiner, \cmp 118(1988)467;  M.A. Shifman and A.V. Turbiner,
\cmp 126(1989)347;
M.A. Shifman, \ijmp A4(1989)2897.}

\lref \hal { M.B. Halpern and E.B. Kiritsis, \mpl A4(1989)1373; A4(1989)1797
(E).}

\lref \haly {M.B. Halpern and   J.P. Yamron,  Nucl.Phys.B332(1990)411;
Nucl.Phys.
B351(1991)333.}
\lref \halp { M.B. Halpern, E.B. Kiritsis, N.A. Obers, M. Porrati and J.P.
Yamron,
\ijmp A5(1990)2275;
  A.Yu. Morozov,  M.A. Shifman and A.V. Turbiner, \ijmp
A5(1990)2953;
A. Giveon, M.B. Halpern, E.B. Kiritsis and  N.A. Obers,
\np B357(1991)655.}
\lref \bpz {A.A. Belavin, A.M. Polyakov and A.B. Zamolodchikov, \np
B241(1984)333. }
\lref \efr {     S. Elitzur, A. Forge and E. Rabinovici, \np B359 (1991)
581;
 G. Mandal, A. Sengupta and S. Wadia, Mod. Phys. Lett. A6(1991)1685. }
\lref \sak {K. Sakai, Kyoto preprint, KUNS-1141-1992. }
\lref \ver {H. Verlinde, \np B337(1990)652.}
\lref \gwz {      K. Bardakci, E. Rabinovici and
B. S\"aring, \np B299(1988)157;
 K. Gawedzki and A. Kupiainen, \pl B215(1988)119;
\np B320(1989)625. }

\lref \sen {A. Sen, preprint TIFR-TH-92-57. }

\lref \bcr {K. Bardakci, M. Crescimanno and E. Rabinovici, \np
B344(1990)344. }
\lref \Jack {I. Jack, D.R.T.  Jones and J. Panvel,  \np B393(1993)95. }
\lref \zam  { Al. B. Zamolodchikov, preprint ITEP 87-89. }

\lref \hor {J. Horne and G. Horowitz, \np B368(1992)444. }
\lref \tse { A.A. Tseytlin, \pl B264(1991)311. }
\lref \GWZNW  { P. Di Vecchia and P. Rossi, \pl  B140(1984)344;
 P. Di Vecchia, B. Durhuus  and J. Petersen, \pl  B144(1984)245.}
\lref \oal { O. Alvarez, \np B238(1984)61. }

\lref \ishi { N. Ishibashi, M.  Li and A. Steif, \prl 67(1991)3336. }
\lref  \kuma  { M. Ro\v cek and E. Verlinde, \np B373(1992)630; A. Kumar,
preprint CERN-TH.6530/92;
 S. Hussan and A. Sen,  preprint  TIFR-TH-92-61;  D. Gershon,
preprint TAUP-2005-92; X. de la Ossa and F. Quevedo, preprint NEIP92-004; E.
Kiritsis, preprint LPTENS-92-29. }

\lref \rocver { A. Giveon and M. Ro\v cek, \np B380(1992)128. }
\lref \frts {E.S. Fradkin and A.A. Tseytlin, \np B261(1985)1. }
\lref \mplt {A.A. Tseytlin, \mpl A6(1991)1721. }
\lref\bn {I. Bars and D. Nemeschansky, \np B348(1991)89.}
\lref \shif { M.A. Shifman, \np B352(1991)87.}
\lref\wittt { E. Witten, \cmp 121(1989)351; G. Moore and N. Seiberg, \pl
B220(1989)422.} \lref \chernsim { E. Guadagnini, M. Martellini and M.
Mintchev, \np B330(1990)575;
L. Alvarez-Gaume, J. Labastida and A. Ramallo, \np B354(1990)103;
G. Giavarini, C.P. Martin and F. Ruiz Ruiz, \np B381(1992)222; preprint
LPTHE-92-42.}
\lref \shifley { H. Leutwyler and M.A. Shifman, \ijmp A7(1992)795. }
\lref \polwig { A.M. Polyakov and P.B. Wiegman, \pl B131(1984)121; \pl
B141(1984)223.  }
\lref \polles { A. Polyakov, in: {\it Fields, Strings and Critical Phenomena},
  Proc. of Les Houches 1988,  eds.  E. Brezin and J. Zinn-Justin
(North-Holland,1990).   }
\lref \kutas {
D. Kutasov, \pl B233(1989)369.} \lref \karabali { D. Karabali, Q-Han Park, H.J.
Schnitzer and
Z. Yang, \pl B216(1989)307;  D. Karabali and H.J. Schnitzer, \np B329(1990)649.
}
\lref \ginq {P. Ginsparg and F. Quevedo,  \np B385(1992)527. }
\lref \gko  {K. Bardakci and M.B. Halpern, \pr D3(1971)2493;
 M.B. Halpern, \pr D4(1971)2398;   P. Goddard,
A. Kent and D. Olive, \pl B152(1985)88; \cmp 103(1986)303;  V. Kac and I.
Todorov, \cmp
102(1985)337.  }
\lref \dvv  { R. Dijkgraaf, H. Verlinde and E. Verlinde, \np B371(1992)269. }
\lref \kniz {  V. Knizhnik and A. Zamolodchikov, \np B247(1984)83. }

\lref \witt { E. Witten, \cmp 92(1984)455.}
\lref \wit { E. Witten, \pr D44(1991)314.}
\lref \anton { I. Antoniadis, C. Bachas, J. Ellis and D.V. Nanopoulos, \pl B211
(1988)393.}
\lref \bsfet {I. Bars and  K. Sfetsos, \pr D46(1992)4510.}
\lref\bsglobal{I. Bars and  K. Sfetsos, \pr D46(1992)4495.}
%

\lref \ts  {A.A. Tseytlin, \pl B268(1991)175. }
\lref \shts {A.S. Schwarz and A.A. Tseytlin, \np B399(1993)691. }

\lref\bsft { I. Bars, preprint USC-91-HEP-B3. }
\lref\BSthree{I. Bars and K. Sfetsos, Mod. Phys. Lett. { A7}(1992)1091.}
\lref\BShet{I. Bars and K. Sfetsos, Phys. Lett. {B277}(1992)269.}
\lref\bs { I. Bars and K. Sfetsos,  Phys. Rev. {D48}(1993)844.}
\lref\bb  { I. Bars, \np B334(1990)125. }
\lref \tsw  { A.A. Tseytlin,\np B399(1993)601.}
\lref \ger { A. Gerasimov, A. Morozov, M. Olshanetsky, A. Marshakov and S.
Shatashvili, \ijmp
A5(1990)2495. }
\lref \hull {C.M. Hull and B. Spence, \np B345(1990)493. }

\lref \sch { K. Schoutens, A. Sevrin and P. van Nieuwenhuizen,
in: Proc. of the Stony Brook Conference {\it `Strings and Symmetries 1991'},
p.558  (World
Scientific, Singapore, 1992).} \lref \boer { J. de Boer and J. Goeree, Utrecht
preprint THU-92/33. }
\lref \dev { C. Destri and H.J. De Vega, \pl B208(1988)255. }
 \lref \bsf { I. Bars and K. Sfetsos, \pl B277(1992)269. }
\lref \kir {{E. Kiritsis, \mpl A6(1991)2871.} }
\lref \nem {D. Nemeschansky and S. Yankielowicz, Phys.Rev.Lett. 54(1985)620;
54(1985)1736(E).}

\lref \br {A. Barut and R. Raczka, ``Theory of Group Representations and
Applications", p.120
 (PWN, Warszawa 1980). }
\lref \jjmo {I. Jack, D.R.T. Jones, N.Mohammedi and H. Osborn, \np
B332(1990)359;
C.M. Hull and B. Spence, \pl B232(1989)204. }

\lref \call {C.G. Callan, D. Friedan, E. Martinec and M.J. Perry,
Nucl. Phys.B262(1985)593; A. Sen, Phys. Rev. D32(1985)316; \prl 55(1985)1846.}
\lref \hul {C.M. Hull, \pl B167(1986)51; Nucl. Phys. B267(1986)266.}
\lref \het { Y. Cai and C.A. Nunez, \np B287(1987)279;
  Y. Kikuchi and  C. Marzban, \pr D35(1987)1400.}
\lref \met {R.R. Metsaev and A.A. Tseytlin, \np B293 (1987) 385.}
\lref \gro {D.J. Gross and J.H. Sloan, \np B291(1987)41. }
\lref  \foa { A.P. Foakes, N. Mohammedi and D.A. Ross, \pl B206(1988)57;
\np B310(1988)335.}
\lref \hw {C.M. Hull and E. Witten, \pl B160(1985)398.}
\lref \hh { D.J. Gross, J.A. Harvey, E. Martinec and R. Rohm, \np B256(1985)253
and  B267(1986)75.}

\lref \chsw { P. Candelas, G. Horowitz, G. Strominger and E. Witten, \np
B258(1985)46. }
\lref \aat { A.A. Tseytlin, preprint CERN-TH.6804/93, hep-th/9302083.}
\lref \at {A.A. Tseytlin, preprint CERN-TH.6820/93. }
\lref \aps { S. De Alwis, J. Polchinski and R. Schimmrigk, \pl B218(1989)449. }
\lref \gny  { M.D. McGuigan, C.R. Nappi and S.A. Yost, \np B375(1992)421. }
\lref \py  {  M.J.  Perry and E. Teo, preprint DAMTP R93/1 (1993); P. Yi,
preprint CALT-68-1852
(1993). }
\lref \tv {   A.A. Tseytlin and C. Vafa, \np B372(1992)443.}
\lref \hull {C.M. Hull and B. Spence, \np B345(1990)493. }
\lref \ell {U. Ellwanger, J. Fuchs and M.G. Schmidt, \pl B203(1988)244; \np
B314(1989)175. }
\lref \ket { S.V. Ketov and O.A. Soloviev, \pl B232(1989)75; \ijmp
A6(1991)2971. }
\lref \GR {A. Giveon and M. Ro\v{c}ek, Nucl. Phys. B380(1992)128.}
\lref \giv {  A. Giveon, Mod. Phys. Lett. A6(1991)2843.}
\lref \GRV {A. Giveon, E. Rabinovici and G. Veneziano, Nucl. Phys.
B322(1989)167;
A. Shapere and F. Wilczek, \np B320(1989)669.}
\lref \GMR {A. Giveon, N. Malkin and E. Rabinovici, Phys. Lett. B238(1990)57.}
\lref \GS  {  A. Giveon and D.-J. Smit. Nucl. Phys.  B349(1991)168.}
\lref  \GP   {  A. Giveon and A. Pasquinucci, Phys. Lett. B294(1992)162.}
\lref \GG  { I.C. Halliday, E. Rabinovici, A. Schwimmer and M. Chanowitz, \np
B268(1986)413. }
\lref \Ve    { K.M. Meissner and G. Veneziano, \pl B267(1991)33;
M. Gasperini, J. Maharana and G. Veneziano, \pl B272(1991)277;
A. Sen,  \pl  B271(1991)295.}
\lref \eng { F. Englert, H. Nicolai and A.N. Schellekens, \np B274(1986)315.}
\lref \ler  {W. Lerche, D. L\"ust and A.N. Schellekens, \np B287(1987)477; \pl
B187(1987)45.}
\lref \fre  { S. Ferrara and P. Fre', \ijmp 5A(1990)989;
M. Billo', P. Fre', L. Girardello and A. Zaffaroni, preprint SISSA/159/92/EP.}
\lref \tye { S. Chung and S.-H. Tye, \pr D47(1993)4546.}
\lref  \GRT   { A. Giveon, E. Rabinovici and A.A. Tseytlin,  preprint
CERN-TH.6872/93,
RI-150-93, hep-th/9304155.}
 \lref \kz {  V.G. Knizhnik and A.B. Zamolodchikov, \np B247(1984)83. }
\lref \sfet { K. Sfetsos, preprint USC-93/HEP-S1, hep-th/9305074.}
\lref  \horne { J.H. Horne and G.T. Horowitz, \np B368(1992)444.}

\lref \nap { M. Henningson and C. Nappi,  preprint IASSNS-HEP-92/88.  }

\lref \givkir {A. Giveon and E. Kiritsis,  preprint CERN-TH.6816/93,
RIP-149-93. }
\lref \kumar { S.K. Kar and A. Kumar, \pl B291(1992)246;
 S. Mahapatra, \mpl A7(1992)2999;
S.K. Kar, S.P. Khastrig and G. Sengupta, \pr D47(1993)3643.
}
\lref \bow  {P. Bowcock, \np B316(1989)80. }
\lref \witten { E. Witten,  Commun. Math. Phys. 144(1992)189. }
\lref \gin  { P. Ginsparg and F. Quevedo, \np B385(1992)527.}
\lref \sft    {K. Sfetsos, \np B389(1993)424.}
\lref \who {
M. Crescimanno, \mpl A7(1992)489;
 I. Bars and K. Sfetsos, \mpl A7(1992)1091; \pl B277 (1992) 269;
  E.S. Fradkin and V.Ya. Linetsky, \pl B277(1992)73;
 A.H. Chamseddine, \pl B275(1992)63.
}

\lref\hor{
P. Horava, \pl { B278}(1992)101.}
\lref \rai {D. Gershon, preprints TAUP-2033-92; TAUP-2121-93; E. Raiten,
``Perturbations of a Stringy
Black Hole'', Fermilab-Pub 91-338-T.
 }

\lref\cres{M. Crescimanno, \mpl A7(1992)489.}
\lref \frlin { E.S. Fradkin and V.Ya. Linetsky, \pl B277(1992)73\hb
 A.H. Chamseddine, \pl B275(1992)63. }

\lref \napwi  { C. Nappi and E. Witten, Phys. Lett. {B293}(1992)309. }
\lref \ST {K. Sfetsos and A.A. Tseytlin, preprint CERN-TH.6962/93,
hep-th/9308018.}
\lref \curci { G. Curci and G. Paffuti, \np B286(1987)399.   }
\lref \ajj {R.W. Allen, I. Jack and D.R.T. Jones, Z. Phys. C41(1988)323. }
\lref \ttt {A.A. Tseytlin, \pl B178(1986)349.}
\lref \hult {C.M. Hull and P.K. Townsend, \np B301(1988)197.}
\lref \jaj {I. Jack and D.R.T. Jones,  \pl B200(1988)453; \np B303(1988)260.}
\lref \tsey {A.A. Tseytlin. \pl B176(1986)92; \np B276(1986)391.}
 \lref \grwi {  D. Gross and E. Witten, \np B277(1986)1. }
\lref\BSslsu{I. Bars and K. Sfetsos, \pl  B301(1993)183.}
\lref\POLWIG { A.M. Polyakov and P.B. Wiegman, \pl  B141(1984)223.  }
\lref \schouten {J.A. Schouten, Ricci Calculus (Springer, Berlin, 1954). }
\lref \busch {T.H. Buscher, \pl B201(1988)466.}
\lref \nem {D. Nemeschansky and S. Yankielowicz, Phys.Rev.Lett. 54(1985)620;
54(1985)1736(E).}
\lref \bpz {A.A. Belavin, A.M. Polyakov and A.B. Zamolodchikov, \np
B241(1984)333. }
\lref \efr {S. Elitzur, A. Forge and E. Rabinovici, \np B359 (1991)
581;
 G. Mandal, A. Sengupta and S. Wadia, Mod. Phys. Lett. A6(1991)1685. }

\lref \bcr {K. Bardakci, M. Crescimanno and E. Rabinovici, \np
B344(1990)344. }
\lref \Jack {I. Jack, D.R.T.  Jones and J. Panvel,  \np B393(1993)95. }
\lref \tse { A.A. Tseytlin, \pl B264(1991)311. }

\lref \frts {E.S. Fradkin and A.A. Tseytlin, Phys.Lett. B158(1985)316;
Nucl.Phys. B261(1985)1. }
\lref \mplt {A.A. Tseytlin, \mpl A6(1991)1721. }
\lref\bn {I. Bars and D. Nemeschansky, \np B348(1991)89.}
\lref \shif { M.A. Shifman, \np B352(1991)87.}
\lref \dvv  { R. Dijkgraaf, H. Verlinde and E. Verlinde, \np B371(1992)269. }
\lref \kniz {  V. Knizhnik and A. Zamolodchikov, \np B247(1984)83. }

\lref \witt { E. Witten, \cmp 92(1984)455.}
\lref \wit { E. Witten, \pr D44(1991)314.}
\lref \anton { I. Antoniadis, C. Bachas, J. Ellis and D.V. Nanopoulos, \pl B211
(1988)393.}
\lref \ts  {A.A. Tseytlin, \pl B268(1991)175. }
\lref \kir {{E. Kiritsis, \mpl A6(1991)2871.} }

\lref \shts {A.S. Schwarz and A.A. Tseytlin,  preprint Imperial/TP/92-93/01
(1992). }
\lref \bush { T.H. Buscher, \pl B201(1988)466.  }

\lref \plwave { D. Amati and C. Klim\v cik, \pl B219(1989)443; G. Horowitz and
A. Steif, \prl 64(1990)260.}

\lref\bsft { I. Bars, preprint USC-91-HEP-B3. }

\lref\bs { I. Bars and K. Sfetsos,  \pr D48(1993)844. }
\lref \tsw  { A.A. Tseytlin,\np B399(1993)601.}
\lref \call {C.G. Callan, D. Friedan, E. Martinec and M.J. Perry,
Nucl. Phys.B262(1985)593.}
\lref \met {R.R.Metsaev and A.A. Tseytlin, Phys.Lett.B191(1987)354;
Nucl.Phys.B293(1987)385.}
\lref \aps { S. De Alwis, J. Polchinski and R. Schimmrigk, \pl B218(1989)449. }
\lref \gny  { M.D. McGuigan, C.R. Nappi and S.A. Yost, \np B375(1992)421. }
\lref \py  {  M.J.  Perry and E. Teo, preprint DAMTP R93/1 (1993); P. Yi,
preprint CALT-68-1852
(1993). }
\lref \witten { E. Witten,  Commun. Math. Phys. 144(1992)189. }
\lref \ajj {R.W. Allen, I. Jack and D.R.T. Jones, Z. Phys. C41(1988)323. }
\lref \ttt {A.A. Tseytlin. \pl B178(1986)349.}
\lref \hul {C. Hull and P.K. Townsend, \pl B191(1987)115;
D. Zanon, \pl B191(1987)363; D.R.T. Jones, \pl B191(1987)363;
S. Ketov, \np B294(1987)813.}
\lref \hult {C.M. Hull and P.K. Townsend, \np B301(1988)197.}
\lref \jaj {I. Jack and D.R.T. Jones,  \pl B200(1988)453; \np B303(1988)260.}
\lref \tsey {A.A. Tseytlin, \pl B176(1986)92; \np B276(1986)391.}
 \lref \grwi {  D. Gross and E. Witten, \np B277(1986)1. }
\lref \myers { R. Myers, Phys.Lett. B199(1987)371;
I. Antoniadis, C. Bachas, J. Ellis and D. Nanopoulos,
\pl B211(1988)393. }
\lref \parall { T.L. Curtright  and C.K. Zachos, \prl 53(1984)1799;
S. Mukhi, \pl B162(1985)345; S. de Alwis, \pl B164(1985)67. }
\lref \tsmpl {A.A. Tseytlin, \mpl A6(1991)1721.}

\lref \calg { C.G. Callan and Z. Gan, \np B272(1986)647. }
\lref \tss {A.A. Tseytlin, \pl B178(1986)34. }
\lref \fri  {D.H. Friedan, \prl 45(1980)1057; Ann. Phys. (NY) 163(1985)318. }
\lref \tsss { A.A. Tseytlin, \ijmp A4(1989)1257. }
\lref \osb { H. Osborn, Ann. Phys. (NY) 200(1990)1.}
\lref \jackk {I. Jack, D.R.T. Jones and D.A. Ross, \np B307(1988)130.}
\lref \jackkk {I. Jack, D.R.T. Jones and N. Mohammedi, \np B332(1990)333.}
\lref \shts  {A.S. Schwarz and A.A. Tseytlin, \np B399(1993)691.}
\lref \GRG {V.V. Zhytnikov, I.G. Obukhova and  S.I. Tertychniy, {\it Computer
Algebra
Program for Gravitation}, in: {\it Abstracts of contributed papers to the
13th Int. Conf. on Gen. Relat. and Grav. (GR-13)} (Huerta Grande, Cordoba,
Argentina, 1992) p. 309.}

\lref \tseyt{A.A. Tseytlin,  preprint CERN-TH.6970/93, hep-th/9308042
(revised).}
\lref \giki{A. Giveon and E. Kiritsis, preprint CERN-TH.6816/93, RI-149-93,
hep-th/9303016 (revised).}
\lref \tsa  {A.A. Tseytlin, \np B294(1987)383. }
\lref \hullt {S. Mukhi, \pl B162(1985)345; \np B264(1986)640; C. Hull and P.K.
Townsend, \pl
B191(1987)115; D. Zanon, \pl B191(1987)363; D.R.T. Jones, \pl B191(1987)363;
S. Ketov, \np B294(1987)813;   M. Bos,
Ann. Phys. 181(1988)177. }
\lref \jacjon {I. Jack and D.R.T. Jones, Liverpool preprint LTP-317 (1993),
hep-th/9310029.}

\newsec {Introduction }

Gauged WZNW theories provided first  examples  of string solutions
which depend non-trivially on $\a'$.
This fact should have   important  implications
 in the context of   establishing  closer
relations between  field theoretic (or \sm) and conformal field theory
approaches.
The derivation of the exact metric $G_{\m\n} $ and dilaton $\p$ corresponding
to the $SL(2,\IR)/\IR$ WZNW model was originally given \dvv\ in the operator
approach which is based on interpreting
the Hamiltonian  $L_0+\bar L_0$ of the coset conformal field theory as a
Klein-Gordon-type operator in a background.
This approach was systematized for establishing the exact form  of
$G_{\m\n}$ and $\p$  in  general $G/H$ coset  models  in \bsfet\sft\BSslsu\aat.

As the group space backgrounds of WZNW models,
the geometries associated with  gauged WZNW  theories in general
have non-trivial  antisymmetric tensor field $B_{\m\n}$
(which is  crucial  for  conformal invariance of the corresponding \sms).
This was  found  in simple  $D=3$ and $D=4$ examples
in the leading  order (`semiclassical') approximation
\horne\hor\rai\gin\napwi.\foot
{There are of course examples of gauged WZNW models (based on non-abelian
groups) with vanishing antisymmetric tensor in the leading order
approximation \cres\BSthree\BShet\frlin.}
To be able to study
the  exact  properties of these geometries  one needs to know the expression
for $B_{\m\n}$ to all orders in $1/k$.

The problem  of  establishing the exact form of the antisymmetric tensor
background turns out  to be quite non-trivial.
The operator approach is not well-suited for
derivation of the  expression for $B_{\m\n}$ since the antisymmetric tensor
does not appear in the zero mode part of the $L_0$-operator.
In principle, one is to consider the $L_0 $  acting on
states $\psi$ of the first excited level and  to  try to deduce the value of
$H_{\m\n\l}=3\del_{[\m}B_{\n\l]}$ by identifying $L_0 \psi$
with the `anomalous dimension operator', i.e
the derivative  $(\del{\bar\b}^i / \del \vp^j)_{\vp_*} $ of the \sm Weyl
anomaly coefficients
parametrized  by the values of  $G_{\m\n}, B_{\m\n}, \p$ at the conformal point
\aat.  This procedure  looks rather indirect and complicated.
Moreover, it  may not work at all beyond the leading order in $\a'$
since the general functional expressions for   the  Weyl anomaly
$\bar\b$-functions
are not known explicitly  and so one is  unable  to  deduce  the exact
expression
for $B_\mn (\a')$  from the comparison of $L_0$ with $(\del{\bar\b}^i / \del
\vp^j)_{\vp_*} $
unless some extra  considerations  (e.g. implying that for some reason there
should exist a scheme in
which $B_\mn$ does not receive $\a'$-corrections at all)  are  invoked.

An  alternative is provided by the effective action approach \tsw\bs\aat\
which, in principle,   offers a direct derivation of the whole \sm action
(i.e.,  $B_{\m\n}$ along with $G_{\m\n}$) at the field-theoretic level.
The  key point is to replace the classical gauged WZNW action $I(g,A)$  by the
quantum effective one $\G(g,A)$,
solve for the $2d$ gauge field and  identify  the local second-derivative
part of the result with the \sm action.
It was proved  in \tsw\bs\aat\ that the operator and effective action
approaches give  identical expressions  for the metric and dilaton.
As for the antisymmetric tensor, there is a subtlety  in its derivation from
the effective action. It turns out that the procedure  of
omitting   the non-local terms in $\G(g,A)$ on the way to a \sm action
{\it a priori } is  not
completely unambiguous in  what concerns the  resulting expression for
$B_{\m\n}$.
Below we shall  consider two
 natural prescriptions for
 the derivation of $B_{\m\n}$ that
preserve the gauge invariance
(present in the full non-local functional $\G (g,A)$)
at the
level of the local part of the effective action.
The gauge invariance is    necessary  in order to be
able to  reduce  (gauge fix)  the \sm from the group  manifold of $G$ to
$G/H$ as a configuration space.
According to the first prescription, $B_\mn$ in general depends on $\a'$,
while $B_\mn$ computed using the second prescription remains semiclassical.

We shall see that in certain cases (when the group $G$  is simple and the
$H$-part
of the matrix
$C_{AB}$  defining the adjoint representation of $G$ is symmetric, or,
equivalently,
  when $\dim H=1$)
the exact expression for $B_{\m\n}$ found using  the first prescription
reduces to the semiclassical
one  (even though the metric and dilaton
still depend on   $\a'$).  For generic simple  $G$ and $H$ the quantum
corrections to $B_{\m\n}$
start with  the `two-loop' $O(1/k^2)$ terms.
We shall present the derivation of the \sm action using  the most
general  expression for the   effective action $\G (g,A)$  which  includes as
particular cases the effective actions for the gauged WZNW model as well as
for the  bosonic \sfet\  and heterotic \ST\
(bosonic part of (1,0) supersymmetric) `chiral gauged' \tye\  WZNW models.

As was  shown  explicitly in the three-loop \ts\ and four-loop \Jack\
approximation in the \sm perturbation theory,
the  exact $D=2$ `black hole' metric-dilaton background of \dvv\ is, in fact,
a solution of the \sm Weyl invariance conditions or the  string effective
equations.
Below we shall  conduct   a similar   check  in the  simplest possible case
with a non-trivial  $B_{\m\n}$:
the three-dimensional   $[SL(2,\IR)\times \IR]/\IR$ (`charged black string')
model \horne.
The exact metric and  dilaton of this model were found in \sft.
As for the antisymmetric tensor, its derivation turns out to be complicated by
the fact that
$G$ is not simple here so that one needs to fix the total derivative ambiguity
in the effective
action in a specific way.

The  Weyl anomaly coefficients ($\bar \b$-functions)  or
the corresponding string effective action are not unambiguous,
i.e. are  scheme (field redefinition)  dependent
\tsey\grwi\met.
The  `two-loop' $O(\a')$-term in the string effective action  depends on a
number of free parameters which change under field redefinitions.
We shall show that there exists such a scheme
(i.e. a choice of the parameters) in which the exact `black string' background
is indeed a solution of the string   equations in the two-loop approximation.
We shall consider in detail several limits of the $D=3$  geometry,
in particular,  the $SL(2,\IR)$
group space and the   direct product of $SL(2,\IR)/\IR$ model and
$\IR$ (`neutral black string'). We shall  discuss the role of the coupling
constant redefinitions  or scheme dependence in  establishing
a  correspondence  between  the conformal
field theory and \sm results.

A particular  limit of the  $[SL(2,\IR)\times \IR]/\IR$ gauged WZNW model was
shown \ST\ to be
equivalent to the $SL(2,\IR)/\IR$ chiral gauged WZNW model
\kumar\sfet\ST.\foot
{It can be shown in general \ST\ that for any abelian subgroup $H$ of a
group $G$ the $G/H$ chiral gauged WZNW model is equivalent to a
$(G \times H)/H$ gauged WZNW model in the axial gauging and  special embedding
of  $H$
 into $G \times H$.}
 Our  derivation of the  exact form  of the \sm couplings in the two
   models gives equivalent results not only    for the
metric and  dilaton  but also for the
antisymmetric tensor.
We shall thus   check  explicitly  that
the  $D=3$ \sm associated with the $SL(2,\IR)/\IR$
chiral gauged model is also conformally invariant at the two-loop level.

In Section 2  we shall find the general expressions for the exact
backgrounds  starting from the effective action for the gauged WZNW model.
We shall consider two natural (`corrected' and `semiclassical')
prescriptions for extracting the \sm couplings  which  give different
expressions for
 the antisymmetric tensor coupling.
 In Section 3 we shall repeat the derivation
using  a more  general  form of the effective action  which
includes as particular  cases the effective actions of the gauged WZNW and
chiral gauged WZNW
models.  The explicit  formulae  for the \sm couplings in Sections 2 and 3 will
be given in the case
when the group  $G$ is simple and a gauged subgroup $H$ is a vector one.

 In Section 4 we shall describe the $D=3$
$\s$-model  originating  from the $[SL(2,\IR) \times \IR]/\IR$ gauged WZNW
model    and show  (in the
$\a'^2$-approximation) that the corresponding exact  background  solves  the
conformal invariance
equations ${\bar \b}^i =0$ in a  proper scheme.  Moreover, we shall find (in
Section 5)
that there exist a scheme in which the semiclassical
limit of this background  is also  a solution of the 2-loop ${\bar \b}^i =0$ -
equations
and will discuss  implications of this fact.

In Appendix A  we  shall prove  that
for a general   model of Section 3
the `measure factor'  $ \exp (-2\p)  \sqrt{\det{G }}$ does not receive
non-trivial quantum corrections. In Appendix B  we  shall discuss the
derivation of  the
exact  couplings  in   a  theory with a non-simple group $G$ -- the
$[SL(2,\IR)\times \IR]/\IR$ model.  We shall show how to resolve the  ambiguity
in the resulting
expression for the antisymmetric tensor coupling  in a way that turns out to be
consistent with
conformal invariance of the $\s$-model. In  Appendix C  we shall  give   the
expressions for some
geometrical quantities used in the computations of Section 4.


\newsec { Exact \sm   corresponding to  gauged WZNW  model}

Let us start by recalling  the basic steps one is to follow in order to
derive the exact \sm
action corresponding to a gauged WZNW or coset  model.
If one solves the classical equations for the $2d$ gauge field $A$,
substitutes the solution back into  the classical WZNW action
$S(g,A)$  \bcr\  and fixes the gauge symmetry,
one obtains a  `semiclassical' \sm  with  the configuration space
$G/H$ which   (with the dilaton coupling included \wit) is
conformally invariant in the one-loop approximation.
To obtain a \sm which is conformally invariant
to all orders in the loop expansion
 one needs (in a `standard' scheme, see Section 5) to modify the \sm couplings
by
$\a'$-dependent terms.
By  definition of the model,  the gauge field $A$ is to be treated as  an
auxiliary field
(for which one does not introduce a source term in the path integral)
but it is not obvious  how to integrate it out completely
(while keeping the group variable $g$ classical)  in a way  that
preserves conformal invariance to all orders.
An approach that preserves conformal invariance is based on first
determining the quantum effective
action $\G (g,A)$  for both $g$ and $A$
(i.e. treating $g$ and $A$ on an equal footing in the path integral)
and then solving for $A$ and eliminating  it from the effective action.
The resulting  gauge invariant functional $\G'(g)$  (restricted to $G/H$)
is to be identified with the
quantum {\it effective}  action  of the corresponding \sm.
That means that the {\it local }  part of $\G'(g)$ should be
equal (after gauge fixing to $G/H$) to the {\it classical}
action  $S(x)$ of the exact \sm  one is  looking for.
It is  clear that in  deriving $S(x)$ from $\G (g,A)$  one is
free to ignore various non-local terms  but  at the same time one must be
careful
to  preserve in the process the gauge invariance that makes
possible to  finally restrict (by the usual procedure of gauge fixing)
the configuration space of the
\sm  to $G/H$.

The classical action of a gauged WZNW model
\eqn\gwznw{\eqalign{ & I(g,A) = I(g)  +{1\over \pi }
 \int d^2 z \Tr \bigl( A\,\bd g g\inv -
 \bar A \,g\inv\del g + g\inv A g \bar A  - A \A \bigr) \ ,\cr
& I \equiv  {1\over 2\pi }\int d^2 z  \Tr (\del g\inv
\bd g )  +  {i\over  12 \pi   } \int d^3 z \Tr ( g\inv dg)^3  \ , \cr } }
is invariant under  the standard  vector $H$-gauge transformations
($A, \A$ take values in the algebra of $H$)
\eqn\inva{ g \ra u\inv g u \ , \  \ A \ra u\inv ( A - \del ) u \ , \ \
 \A \ra u\inv ( \A - \bd ) u \ , \ \  \ \ u = u (z, \bar z)\in H  \ . }
In this section  we shall assume
that the group $G$  and the subgroup $H$  are simple.
 Parametrizing  $A$ and $\A$ in terms of
subgroup elements
 $h$, $\bh \in H$
which transform as $ h \ra u\inv h \ , \ \ \bh \ra   u\inv \bh $,
\eqn\para{ A =\del h h\inv \ , \qq  \A = \bd \bh \bh\inv  \ ,}
one can use  the  Polyakov-Wiegman identity \POLWIG\ to represent the
action  \gwznw\ in terms of the two WZNW actions corresponding to the group $G$
and   to the  subgroup $H$,
\eqn\gwzwh{\eqalign{ &I(g,A)= I({\tilde g} ) - I({\tilde h} )  \ ,\cr
&{\tilde g} = h\inv g \bh \ , \qq {\tilde h}  = h\inv\bh\ . \cr } }
The corresponding quantum effective action \tsw\bs\  has a simple form in
terms of ${\tilde g} ,   \  {\tilde h}$
(we ignore extra non-localities introduced by field renormalizations
because they give rise to non-local terms in the $\s$-model, see  \aat)
\eqn\eff{ \G (g,A) = (\kg) I(h\inv g \bh) -  (\kh )I(h\inv \bh)     =
 (\kg) \bigl[ I (g, A) \  - \ b  \ \O (A)  \bigr]\ , }
where $\gG$, $\gH$ are the dual Coxeter numbers for the group $G$ and the
subgroup $H$ respectively and $ b \equiv  - { \gG - \gH \over \kg } $.
$\O (A)$  is the non-local  gauge invariant functional of $A $, $  \A $,
\eqn\funct{ \O (A) \equiv I(h\inv \bh)
= \o (A) + \W (\A) + \1p \int d^2 z \Tr (A\A)\ ,}
where the non-local functionals $\o(A)$ and $\W(\A)$ are defined as
\eqn\fun{\eqalign{ & \o (A)  \equiv  I(h\inv) = - \1p \int d^2 z
\Tr \bigl \{ \half A{\bd\over \del} A
+ \third A[{1\over \del}A , {\bd\over \del}A] +  O(A^4)  \bigr \}\ , \cr
& \W(\A)  \equiv   I(\bh)
 = -\1p \int d^2 z \Tr \bigl \{\half \A{\del\over \bd} \A
- \third \A[{1\over \bd}\A, {\del\over \bd}\A] + O(\A^4) \bigr \} \  .\cr } }
Before solving the equations for $A$ and $\A$, which follow from \eff,
let us first  drop the non-local cubic and higher order in $A,\A$ terms in
\fun.
As explained above, the non-local terms should not contribute to the \sm
action which is our final goal.
The truncated action has the form
\eqn\gtr{ \G_{tr}(g, A) = (\kg) \bigl[ I (g, A )  - b\ \O_{tr} (A) \bigr]  \ ,}
where
%
\eqn\otr{ \O_{tr} (A) \equiv  {1\ov 2\pi} \int d^2 z \Tr (A-\tbA)(\A-\tA) \ ,}
or
$$ \O_{tr} (A) =  \2p \int  \Tr  F_0 {1 \over \del \bd } F_0  \ , \  \ \ F_0
\equiv \bd A
- \del \A \ .    $$
The fields   $\tA$, $\tbA$  in  \gtr\  are defined by
\eqn\tatba{ \tA \equiv {\bd \ov \del} A =\bd h h\inv +\dots \ ,\qq
\tbA \equiv {\del \ov \bd} \A = \del \bh\bh\inv +\dots \ , }
and transform under the gauge transformations as
\eqn\tra{ \tA \to u\inv (\tA-\bd) u\  + \dots ,\qq \tbA \to u\inv (\tbA - \del)
u  + \dots \ .}
Here (and in  similar equations below)   dots  stand for  contributions of
higher order non-local terms.
Note that in \funct\ the  Polyakov-Wiegmann formula was used, which presumes
an integration by parts.
In \gtr\ we have undone this integration by parts, thus restoring the
manifestly gauge invariant form of the quantum term under the
transformation of the gauge fields \inva\ and \tra.
 One may question whether the  truncation  of \eff\ to \gtr\ is legitimate
since
$O(A^n,\A^m)   $ terms with $  n,m \geq 3\  $  in \eff\ are necessary for
gauge invariance.
In fact,  \otr\ is invariant only under the `abelian' part of the gauge
transformations.
The point, however, is that since  the non-abelian gauge invariance  is
violated only  by
{\it non-local}  terms (or, equivalently, is  preserved  up to  {\it non-local}
 terms)  it   should
be present in a   consistently extracted  local  part of \gtr.

To proceed,   let us first   establish our notation.
$T_A = (T_a , T_i )$ are  the
generators  of $G$;  $\ T_a$ are the generators of  $H$; $\ A=
1,\dots, D_G ; \ a=1,\dots, D_H; \ i=1,\dots, D, \ D=D_{G/H}; \
 \eta_{AB}$ is negative definite in the compact case,  and
\eqn\notation{\eqalign{ & A= A^aT_a\ , \qq
C_{AB} \equiv  \Tr ( T_A g T_B g\inv ) \ , \qq
\Tr (T_AT_B) = \eta_{AB} \ ,\cr
&  J_A =  \Tr (T_A g\inv\del g)=  E_{AM }(x)  \del x^M \ ,\qq
 \J_A =- \Tr (T_A \bd g g\inv )= \E_{AM }(x)  \bd x^M \ , \cr
&{\tilde J}_A =  \Tr (T_A g\inv\bd g)=  E_{AM }(x)  \bd x^M \ , \qq
{\tilde {\J }}_A =-\Tr (T_A \del g g\inv )= \E_{AM }(x)  \del x^M \ ,\cr
& \E_{AM}=-C_{AB} E^B_M\ ,\qq  C^{AD}  C_{BD}= \delta^A_B \ , \qq  C^A_{\ B } =
\eta^{AD} C_{DB} \ ,
\cr  & C_{AB} =     \left(\matrix { C_{ab}    &     C_{ib} \cr
   C_{ aj}  &   C_{ij}    \cr}\right)   \ ,  \qq \E_{aM} = - C_{ab} E^b_M -
C_{ai}
E^i_M \ .  } }
In the following we shall also use\foot{Our present notation differ from
the notation of ref. \aat\  where $A,\A, \J$ and $\E$ have  the opposite signs.
The correspondence with the notation of  ref. \bs\ is the following:
 $A_- \ra A, \ A_+ \ra  \A, \ (L^A_M, R^A_M ) \ra (E^A_M, \E^A_M )$. }
\eqn\defs{A= (A^a)\ , \qq  J= (J_a) \ ,\qq
 M_{ab} \equiv C_{ab} - \eta_{ab} \ ,\qq N_{ab}\equiv M_{ab}-b\ \eta_{ab}\ .}
Then the truncated action \gtr\ takes the form
\eqn\trun{\eqalign{ & \G_{tr}(g,A)=(\kg) \bigl[ I(g) + \Delta I (g,A) \bigr]\ ,
 \cr
&\D I (g, A)\equiv {1\over \pi } \int d^2 z \  \big[  - A\J  -
 \bar A J   +    A  M \A
-  \ha  b  \ ( A- \tbA) (\A- \tA)   \big]  \ . } }
Eliminating $A,\A$ from this action one finds a non-local expression
which is bilinear in $J,\J$
$$   \D I (g) =-{1\over 2 \pi }
 \int d^2 z \ \big[ J (QN^TQ\inv N - b^2)\inv (QN^T Q\inv \J - b  QJ) $$ $$
+ \J (Q\inv NQ N^T - b^2)\inv (Q\inv N QJ -bQ\inv \J)\big]   \ ,
\ \ \ \ \  Q\equiv  {\bd\over \del} \ , \ \  Q^{-1}\equiv  {\del \over \bd} \ .
\
\eq{2.14'} $$
One is now to  take  a local   part of the functional $ I(g) + \D I (g)$  and
identify it with the
\sm action.  To determine the metric term in the resulting \sm action it is
sufficient just to set
$Q=1$ or to take the $d=1$ limit of the action \bs\aat. As for the  parity-odd
(antisymmetric tensor)
part in the  \sm action its derivation is more subtle. It is not clear a priori
 how to extract
the relevant local part of (2.14$'$). One natural suggestion \bs\aat\ is to
replace $ QN^TQ\inv$  and
$ Q\inv NQ$ by
$N^T$ and  $N$  and  $QJ$ and $Q\inv \J$ by $\tilde J$ and ${\tilde {\J }}$.
The resulting local action does not, however, have gauge invariance in its
parity-odd part.
As we shall show below, there exist a procedure
of extracting a local
part from (2.14$'$)  that not only  preserves the  abelian gauge invariance of
(2.14), (2.14$'$)
but also  restores the full non-abelian invariance  which was present in  the
original action \eff.
The basic idea  is to omit the non-local terms  already at the level of the
solution of the
equations for $A, \A$ that follow from (2.14) and only then substitute the
resulting local
expressions for $A,\A$ $and$ $\tA,\tbA$ into (2.14).

The  equations for $A^a, \A^a$ we obtain from \trun\  are
\eqn\eqm{ -\J  + N\A + b\ \tA  = 0 \ ,\qq
-J + N^TA  +b\ \tbA  =0 \ .}
Let us note that \eqm\ take the form of the exact equations that follow from
the full untruncated
effective action \eff\ if one sets there (cf. eq. \tatba)
$\tA  =\bd h h\inv , \ \tbA = \del \bh\bh\inv .$
Eqs. \eqm\  imply
\eqn\eqmc{ -\tbJ  + N\tbA + b\ A + \dots  = 0 \ , \qq
- \tJ + N^T\tA   + b\ \A+ \dots =0 \ . }
Up to non-local terms the  solution of \eqm\ and \eqmc\ is
(we  shall use the star to denote the local part of the solution)
\eqn\sol{\eqalign{ & A_* = V\inv(N J  - b\tbJ) \ , \qq
\A_*= \V\inv ( N^T \J - b  \tilde J)  \ , \cr
&\tA_* = V\inv(N \tJ  - b \J) \ ,\qq
 \tbA_*= \V\inv ( N^T \tbJ - b   J)\ ,\cr }      }
where we have defined the matrices
\eqn\nvv{\eqalign{& V\equiv N N^T - b^2 I\ , \qq \V\equiv N^T N - b^2 I  \ ,\cr
&V^T= V \ , \qq \V^T= \V \ , \qq   \V\inv N^T = N^T V\inv \ .\cr }  }
The local  expressions   in \sol\ have the correct Lorentz  structure.
Moreover, $A_*$, $\A_*$ and $\tA_*$, $\tbA_*$ transform properly
(as gauge  fields) under the {\it non-abelian}  gauge
transformation of $g$.
In fact, the gauge transformation $g \ra u\inv g u $  induces the following
transformations on the $H$-components of the currents and tensors
\eqn\tran{\eqalign{ &J' = U J  - {M'}^T \ep \ , \qq  \J'= U\J - {M'}\be \ , \qq
 \tJ' = U \tJ  - {M'}^T \be \ , \qq \tbJ'= U\tbJ - {M'}\ep \ ,\cr
&  C'\equiv  C(g')= UC(g)U\inv \ ,\qq M' = UMU\inv \ ,\qq  V' =  UVU\inv \ ,\qq
\V' =  U\V U\inv \ ,\cr
&U_{ab} \equiv C_{ab} (u\inv ) =  C^T_{ab} (u) \ , \qq U^T=U\inv \ , \qq
 u\inv T_a u = C_{ab} (u) T^b\ ,\cr
&\ep_a \equiv \Tr (T_a u\inv \del u )\ , \qq
\be_a \equiv \Tr (T_a u\inv \bd u )  \ . }   }
Using \tran\  one can show  that  the fields in \sol\
transform as
\eqn\trans{
A_*'= U A_* - \ep \ , \qq  \A_*'= U \A_* - \be \ ,  \qq
\tA_*'= U \tA_* - \be \ , \qq  \tbA_*'= U \tbA_* - \ep \ ,      }
i.e. as the gauge fields in \inva, \tra.  Note also that
$$ (A_*- \tbA_*)' = U(A_*- \tbA_*)\ , \ \  (\tA_*- \A_*)' = U(\tA_*- \A_*)\ ,
$$ $$  \ \
  [(A_*- \tbA_*)^a(\tA_*- \A_*)_a]' = (A_*- \tbA_*)^a(\tA_*- \A_*)_a\  . $$
Therefore,  substituting  \sol\ into \trun\ we  can
  eliminate  the gauge fields  in a way that preserves
gauge invariance.

A natural  question is  whether  the  described  procedure  (which we shall
call `corrected'
prescription)  based on \sol\ is a unique one preserving gauge invariance.  In
fact, it is not --
there exists  another similar prescription
(which will be  described at the end of this section and will be called
`semiclassical') in
which the  local part of the quantum correction  (2.14$'$) does not contain a
parity-odd part  and
thus the expression for $B_\mn$ is not modified. To see that the prescription
using \sol\ is not
unambiguous  note that we have dropped  non-local terms in the process of
solving \eqm\
(the  last two relations  in \sol\ imply  also that we  assume a specific
prescription
of separating out local terms in  (2.14$'$)).
 Eliminating $A,\A$ from  the action \eff, \trun\  one  has to  be  careful
about doing
 integrations by parts: total derivative  terms  of  non-local structure
may,  in fact, contribute to the local part of the result (disregarding  total
derivative terms in the
action does not commute with  dropping  out   non-local terms after the
insertion  of the
`truncated' classical solution  into the action).
For example, let us consider the effect of adding a total derivative term
$$ -\ha \mu b  (A\A - \tA\tbA)  $$
to the action (2.14) ($\mu$ is an arbitrary parameter). Such a term will not
change the equations of motion (2.15) but will produce a non-trivial
contribution to
the antisymmetric tensor coupling after  the solution (2.17) is substituted
back into the action
(the expression for the metric will not change).  It should be  possible to
modify
the above  term by other $g$-dependent total derivative terms
(which should compensate for the use of the Polyakov--Wiegmann relation in the
process of
representing the manifestly gauge invariant  action \eff\ in the
standard  `$AJ$'-form (2.14))
in order to
preserve the gauge invariance of the result.
It turns out to be necessary to account for this ambiguity in the case of a
non-simple group $G$
(see Appendix B).  For a simple $G$ adding such  terms  seems  unnatural since
 they   change the coefficient of the $A\A$-term in the local part of (2.14).

 Starting with (2.14), we find the  following local action for $g$
\eqn\trloc{\G_{tr}(g, A) = (\kg) \bigl[ I (g) +  \D I (g, A_*)  + \dots \bigr]
 =  \G_{loc}(g) + \dots \ ,}
where\foot
{
In the treatment of the general model in \bs\aat\
gauge invariance was maintained only  in the zero mode sector.
As a result, the expression  for
the  antisymmetric tensor we find below (2.25)  is different from  the one
given in \bs\aat.
  The  form  of the truncated effective action used in \bs\aat\  was  different
from
\gtr\  by  the  total derivative term (with $\mu=1$) and as a result
a gauge-variant expression  for the $B_{MN}$-term  was obtained
 after separation of the local part of the action.  }
\eqn\local{\eqalign{ \G_{loc}(g) & =(\kg)\bigl\{ I(g)+{1\ov 2\pi}
\int d^2z \ \bigl[ -A_* \J - J\A_* +b(A_*\A_* -\tbA_*\tA_*) \bigr] \bigr\} \cr
&= (\kg) \bigl\{ I (g) +  {1\over  2\pi }
 \int d^2 z \ \bigl( \  J\ [-2\V\inv N^T + b \V\inv (N^{T}N^T - b^2 I) V\inv ]
\ \J \cr
 & - \  b \tbJ\ V\inv  (NN-b^2 I) \V\inv \tJ
+ \  b J \ [ \V\inv   + b \V\inv(N-N^T)\V\inv ]\  \tJ \cr
& + \ b \tbJ\ [ V\inv   +  b V\inv(N-N^T)V\inv ]\  \J \ \bigr) \bigr\}\ . }}
As follows from the above discussion, this action is
invariant  under the transformation $g\to u\inv g u$.
Parametrizing $g$ in terms of group coordinates $x^M$ (see \notation) we  can
represent  \local\
as the following  \sm with the group space $G$ as a configuration space\foot{We
assume that the
coordinates $x^M$ are rescaled by the overall `radius' of the space  so that
$\a'$ is dimensionless.}
\eqn\sloc{S(x)= \G_{loc}(g)=
- {1 \over \pi \a' } \int d^2 z \ {\cal G}_{MN} (x) \del x^M \bd x^N \ , \qq
\a' = {2\ov \kg } \ ,}
where the metric is
\eqn\metric{ G_{MN} \equiv  {\cal G}_{(MN )} = { G}_{0MN }  +
 2 (  \V^{-1 } N^T )_{ab} E^a_{(M} \E^b_{N)}
 - b  (\V^{-1 })_{ab} E^a_M E^b_N
-  b (V^{-1 })_{ab} \E^a_M \E^b_N \ , }
and the antisymmetric tensor is
\eqn\antis{\eqalign{  B_{MN} \equiv  &{\cal G}_{[MN ]} = B_{0MN} +
2\bigl[\V\inv N^T - b \V\inv (N^{T}N^T - b^2 I) V\inv \bigr]_{ab}
E^a_{[M}\E^b_{N]} \cr
&- b^2 \bigl[ \V\inv(M-M^T)\V\inv \bigr]_{ab}  E^a_{[M} E^b_{N]}
-  b^2 \bigl[ V\inv(M-M^T)V\inv \bigr]_{ab}  \E^a_{[M} \E^b_{N]} \ .\cr }}
 ${ G}_{0MN }$  and ${ B}_{0MN }$  are  the  original  WZNW
(group space) couplings,
\eqn\orc{ G_{0MN} =\eta_{AB} E^A_M  E^B_N  \ , \qq
 3\del_{[K} B_{0MN]} =  E^A_KE^B_ME^C_N f_{ABC} \ .}
%
The expression for the dilaton coupling  follows from the (local part of)
determinant of the $(A,\A)$ quadratic form  in \trun
\eqn\dila{ \p = \p_0  - \fourth  {\ \rm ln \ det \ }  V   \ . }
It is useful to rewrite the expressions \metric, \antis\
separating quantum corrections from the semiclassical expressions
\eqn\semicl{ { G}_{MN }^{(s)} ={ G}_{0MN }  +
 2 M\inv_{ab} E^a_{(M} \E^b_{N)} \ , \qq    { B}_{MN }^{(s)}=B_{0MN} +
  2M^{-1 }_{ab}   E^a_{[M} \E^b_{N]} \ . }
Namely,
\eqn\metri{ G_{MN}  = { G}_{MN }^{(s)}
+  2 b (  \V^{-1 } M^TM\inv )_{ab} E^a_{(M} \E^b_{N)}
 -   b (\V^{-1 })_{ab} E^a_M E^b_N -   b (V^{-1 })_{ab} \E^a_M \E^b_N \ ,}
and
\eqn\ants{\eqalign{B_{MN}  =&{ B}_{MN }^{(s)} +  \
  2b^2 \bigl[  \V^{-1 } M^TM\inv(M-M^T)V\inv   \bigr]_{ab} E^a_{[M} \E^b_{N]}
\cr
&  - b^2 \bigl[ \V\inv(M-M^T)\V\inv \bigr]_{ab}  E^a_{[M} E^b_{N]}
-  b^2 \bigl[ V\inv(M-M^T)V\inv \bigr]_{ab}  \E^a_{[M} \E^b_{N]}   \  . } }
Thus the first quantum correction to  the metric $G_{MN}$  is one-loop
($O(\a')$)
and to $B_{MN}$ --  a two-loop
($O(\a'^2)$) one. Let us note that the  above expressions  were derived in the
case of the vector
gauging of a simple group $G$.

  The \sm action \sloc\ is  gauge invariant as a
consequence of the   invariance of \local\ (note that the semiclassical and
quantum correction  terms
in \local\ and \sloc\ are separately gauge invariant).
The general \sm \sloc\ is invariant under the  {\it local}  transformation $
\delta x^M= Z^M_a (x)
\epsilon^a (z,\bar z)$   (e.g. induced by  the gauge transformation
$g' = u\inv g u$   with  generators   $Z^M_a
= E^M_a + \E^M_a $) if  the metric and the antisymmetric tensor satisfy the
following
constraints\foot{Since we are deriving the constraints for the couplings $G,B$
(which transform only
as functions of $x$)
 the variation of the action should vanish for arbitrary $x$ and $\ep$.
In the metric part of the variation  one
should equate to zero separately the $O(\ep)$ and $O(\del \ep)$ terms
(integration by parts
in $O(\del \ep)$ terms  induces $\del\bd x$-terms). }
 \eqn\zettr{ Z^K_a \del_K G_{MN} + G_{MK} \del_N Z^K_a  + G_{KN} \del_M  Z^K_a
 =0 \ , \ \ \ G_{MN} Z^M_a =0 \ , \ \ \  H_{MNK} Z^K_a =0 \ , }
so that  $Z^M_a$ should be the Killing vectors as well as the null vectors of
the  metric
  and  the antisymmetric tensor field strength  should have a zero projection
on $Z$ (cf. \jjmo).
These conditions  are thus  satisfied for the metric \metric\ (see \aat) and
the antisymmetric tensor \antis. The dilaton term is also invariant ($\p(g') =
\p (g) $
or  $Z^M_a \del_M\p =0$)   as it is clear from \dila, \tran.

 One can now  fix a gauge, e.g.  by
restricting the  coordinates  $x^M$ on $G$ to coordinates  $x^\m$ on $G/H$.
This can be done by
solving a gauge condition $R^s (x^M) =0$ ($s=1,\dots, \dim H, \ $  $ \det
(Z^M_b \del_M R^s)  \not= 0
$),  i.e.   $x^M=x^M(x^\m) $ ($\m = 1,\dots, \dim (G/H)$)  and  $ \del_M R^s
\del_\m x^M =0$.    In
particular, it is possible to choose   such $x^\m$  that are invariant under
the transformations
of $x^M$, i.e  $Z^M_a \del_M x^\m=0$.\foot {Since
 the  action \sloc\   has  gauge invariance it
really depends on the $\dim (G/H)$ invariants $x^\m$ (global  coordinates)
 that one can build  out of
$\dim G$ group parameters $x^M$ \bsglobal.
Such choice  has certain advantages
over  a generic  procedure of gauge fixing. Fixing  a  gauge one   usually
restricts  consideration  to  one  patch of the entire  space only, whereas
choosing  the
$G/H$  coordinates $x^{\m}$ as group invariants
one may  determine their range
of values  by using group theoretic methods \bsglobal.}
 If $H^M_i$ is a  basis in the tangent space of
$G/H$ orthogonal (with respect to $G_{0MN}$)
to the  gauge symmetry generators $Z^M_a$ then, in particular, we can choose
 $\del_\m
x^M \equiv
 H^M_\m =  L^i_\m
H^M_i$  so that
 the couplings of the resulting \sm on $G/H$  are given by
\eqn\gaug {G_{\m\n}    = G_{MN}H^M_\m H^N_\n = {\bar g}_{ij} \bH^i_\m \bH^j_\n
\ ,
\   \ \  {\bar g}_{ij} = \eta_{ij} -   b  V^{-1}_{ab} C^a_{\ i} C^b_{\ j}\  , }
\eqn\antg{  B_{\m\n}    = B_{MN}H^M_\m H^N_\n
=
{\bar b}_{ij} \bH^i_\m \bH^j_\n \ , \ \
{\bar b}_{ij} = b_{0ij} -  b^2 \{ V\inv(M-M^T)V\inv \}_{ab} C^a_{\ i} C^b_{\
j}\ , }
  $$  \ H_{\m\n\l } =
{\bar h}_{ijk} \bH^i_\m \bH^j_\n \bH^k_\l \  .   $$
Here $\bH^i_\m = \bH^i_M\del_\m x^M$, with   $\ \bH^i_M$ being  a particular
basis
orthogonal to  $Z^M_a$  \aat\
\eqn\bas{  \bH^i_M = E^i_M - M^{-1}_{ab} C^{bi} E^a_M\ ,  }
and  $b_{0ij}$ is the projection of $B_{0MN}$.
 Using that
\eqn\meta{C_{ai}(u)=0 \ , \qq uT^i u\inv =T^jC_j{}^i(u)\ ,\qq
C_i{}^k(u) C_{jk}(u)={\eta}_{ij} \ ,}
one can  show that under  $g\to u\inv g u$
the 1-form ${\bar H}^i=\bH^i_M dx^M$ and the tensors
${\bar g}_{ij}$, ${\bar h}_{ijk}$ transform as
\eqn\traans{\bH^i{}' = \bH^j C_j{}^i(u)\ , \ \
{\bar g}'_{ij} = C^k{}_i(u) {\bar g}_{kl} C^l{}_j(u) \ ,\ \
{\bar h}'_{ijk} =  {\bar h}_{qpn}C^q{}_i(u)C^p{}_j(u) C^n{}_k(u) \ .}
As it is clear from \metri,\ \ants,  in the case when $C_{ab}$ (and hence
$M_{ab} $ and $ N_{ab}$)
is  symmetric  the metric still receives $1/k$ - corrections  while the
antisymmetric tensor remains
semiclassical
 \eqn\symm{ G_{MN}    = { G}_{MN }^{(s)}
  - b  [M(M-2b)]^{-1}_{ab} (E^a_M  - \E^a_M )(E^b_N  - \E^b_N)  \  ,
\ \ \  B_{MN} =  { B}_{MN }^{(s)}   \ .}
It is possible to  prove that the
 matrix $C_{ab}$ is symmetric  only when  the subgroup $H$ is  abelian
with $\dim H=1$.  Expanding the group element $g= \exp (X^AT_A) $
  around the identity
one  can  compute the  expansion of $C_{ab}$ in \notation\ to $X^3$-order.
Demanding  that the $O(X)$ term  is
symmetric  we obtain  the condition
$f_{abc}  X^c =0.$  For this to be true for all  $X^c$
the structure constants $f_{ab}{}^c$ of the
subgroup $H$  should  vanish   so that
 $H$  should be  abelian.
%
Assuming that $H$ is abelian we find  an additional condition from the $O(X^3)$
term:
 $ \ X^i f_{ai}{}^j \Tr\bigl[\ T_j T_b (X_C T^C)^2 \ \bigr]
- (a \leftrightarrow b) =0  .$
This  condition can be satisfied  for all $X$'s only if $\dim H=1$.  One can
also see this
explicitly  using  the general form of the expansion  of $C_{AB}$
in normal coordinates $X^M$ near the unit element of the group (see e.g.
\schouten)
\eqn\norm{E^A_M = \big({\e{f} - 1 \ov f } \big)^A_M = (I + \ha f + {1\ov 6} f^2
+ \dots )^A_M \ , \
\  \ f_{AB}\equiv f_{ABC}X^C \ , \ \ \ E^A_MX^M=X^A\ , }
 \eqn\expan{
C^A_{\ B} = ({\rm e}^{-f})^A_{\ B} = (I - f + \ha f^2 - {1\ov 6}   f^3 +\dots
)^A_{\ B}  \  . }
Imposing the symmetry condition on the $O(X)$ and $O(X^3)$ terms in  $C_{ab}$
one
concludes that  $\dim H=1$.

In the case of a simple group $G$ considered in this section  the condition
that $C_{ab}$ should be
symmetric or that  $\dim H=1$ is a  sufficient  one in order for the quantum
corrections in \ants\  to be absent.  The matrix  $C_{ab} $ is not  symmetric
in a
generic  case of abelian or  non-abelian  $H$.

  The situation is different when   $G$  is not simple.  The general
expressions for the
effective action \eff, \local\ and hence for the \sm couplings in \sloc\
 become more complicated but  can be  worked out by taking into account that
the renormalizations
of the levels $k$  may  be different for  different  simple factors in $G$.
It turns out also to be necessary to  fix the `total derivative' ambiguity in
the
derivation of the expression for  the antisymmetric tensor in a particular way,
consistent with conformal invariance of the resulting \sm.
 One of the consequences is that  the antisymmetric tensor (computed according
to our first
prescription based on \sol)
 may
contain  quantum corrections even when
$\dim \ H = 1 $. An example of such model will be discussed in Section 4  and
Appendix B.
Another  example  is provided by the $SO(2,2)/SO(2,1)$ model
(with non-symmetric
$C_{ab}$ \bs\ and vanishing semiclassical $B_{\m\n}$ \cres\BSthree\frlin)
for which we have  numerically checked
(in a certain gauge) that the  quantum correction to the  antisymmetric
tensor  is also vanishing (the $O(E\E)$ term in \ants\ vanishes by itself while
the  contributions of
the  $O(EE)$ and $O(\E\E)$ terms cancel each other).
The matrix  $C_{ab} $ is non-symmetric  also in the case of the $D=4$ model
($SL(2,\IR)\times SU(2)/ \IR\times U(1)$) considered in \hor\napwi\BSslsu\  but
here  it appears
that  the antisymmetric tensor receives non-trivial corrections to all orders
in $1/k$.

Let us now describe another natural prescription of separating  a local
gauge-invariant  part
of the quantum term  in  the effective action (2.14$'$) under which $B_\mn$
does not receive
$\a'$-corrections.
Suppose that one has  managed to truncate (2.14$'$) in such a way
that the resulting local part (quantum contributions to it)
is $2d$ parity-even
(invariant under $z\ra {\bar z}, \ {\bar z}\ra  z$),
i.e. does not contain any $\a'$-corrections to the  antisymmetric tensor part.
That would mean that
the  resulting metric is still given by (2.24) but $B_\mn$ remains
semiclassical as in (2.28) (however, with  overall coefficient
$(k+{\rm g_G})$).  The
corresponding  \sm is obviously gauge invariant since the semiclassical  part
of (2.14) is gauge
inavriant by itself    and the parity-even and parity-odd parts  of  (2.14)
should be gauge invariant
separately (the gauge transformations (2.19), (2.20) do not mix parity-odd and
parity-even sectors).

An indication that such a result may be considered as a natural one comes from
the structure of the
quantum term in the truncated effective action (2.8). This term is manifestly
parity-even, suggesting
that  there should be no correction to the parity-odd part of the \sm that is
obtained after
solving for $A,\A$. The fact that the quantum term (truncated to $O(A^2)$ part)
in the effective
action (2.8) is parity-even may be considered as a basic `principle' behind
this prescription.
This may be in turn related to the fact that from the point of view of the
operator conformal
field theory approach  it is the Hamiltonian but not the current algebra (which
is sensitive to
a definition of currents and thus to the form of $B_\mn$) that receives $1/k$
corrections.

 It is easy indeed
to give  a simple prescription  for  solving for $A,\A$  (up to non-local
terms) in (2.8) that
does not generate parity-odd quantum terms in the \sm action and thus
corroborates this suggestion.
It is useful to separate the classical term  $A_{cm}$ in the solution for $A_m$
from
the quantum one $A_{qm}= O(b)$ (see (2.14))
$$  A= A_c + A_q \ ,  \ \ \ \A= \A_c + \A_q \ , \ \ \ A_c = M\inv J\ ,  \ \ \
\A_c = M\inv \J\ .
\eq{2.40} $$
Then (2.14) takes the folowing form (cf.(2.22))
$$ \G_{tr}(g,A_q)=(\kg) \bigl[ I_c (g)
 + {1\over \pi } \int d^2 z \  \bigl(     A_q  M \A_q  $$ $$
-  \ha  b  \ [ A_q- \tbA_q + M\inv (J- \tbJ)] [\A_q- \tA_q + M\inv (\J - \tJ)]
  \bigr)  \ ,
\eq{2.41}   $$
where $I_c$ is the semiclassical term
$$ I_c (g) =  I(g) -   {1\over \pi } \int d^2 z \  JM\inv \J \ . \eq{2.42} $$
In (2.41) we have already dropped some non-local terms  by using the rule
(similar to that used in \sol): in substituting the classical fields $A_c,\A_c$
into the quantum
term in (2.14)  we replaced $ \tA_c = {\bd \ov \del} ( M\inv J ) $ by   $ M\inv
\tJ$
and $ \tbA_c = {\del \ov \bd} ( M\inv \J ) $ by   $ M\inv \tbJ$  (see
(2.10),(2.12)). The  second  part of our prescription is to  maintain  the
parity-even structure of the quantum part of (2.41)  by symmetrising
the $ A_q  M \A_q$-term,
$$  A_q  M \A_q  \ \ \ra \ \   \ha   (A_q  M \A_q  + \tA_q M \tbA_q)  \ .
\eq{2.43} $$
This is equivalent to dropping   the term
$  \ha   (A_q  M \A_q  - \tA_q M \tbA_q) $   which is a total derivative up to
non-local terms depending on derivatives of $M$ (cf. a  discussion  above
eq.(2.21)).
The resulting expression for the quantum term $\G_{tr}(g,A_q)- (\kg)  I_c (g)$
is  parity-even  (and  gauge-inavriant). After  one solves for $A_q, \A_q$ one
finds, therefore,   a
\sm with no quantum  correction to the antisymmetric tensor term, i.e.  with
the metric and  dilaton
 given by  (2.24) and (2.27) and the semiclassical antisymmetric tensor (2.28).
It should be kept in mind of course that it is still the shifted level $\kg$
that appears in front
of the semiclassical  $B_\mn$ in the \sm action.

As we shall check below   on the example of  a $D=3$  gauged WZNW model   (see
Sections 4, 5 and
Appendix B) both    backgrounds   derived  according to the two (`corrected'
and `semiclassical')
prescriptions for $B_\mn$ described in this section   are
 solutions of the \sm conformal invariance conditions in the two-loop
approximation.


\newsec{Exact $\s$-models  corresponding to   bosonic
and heterotic chiral gauged WZNW models}

Using the equivalence of the (1,1) supersymmetric gauged WZNW models with the
$N=1$ superconformal coset models it was shown in \bsfet\Jack\ that the
expressions
for the bosonic background fields remain semiclassical.
The (local part of) the effective action of the (1,1)
supersymmetric gauged
WZNW model is also given just by the classical gauged WZNW action itself \aat.
The  (1,0) supersymmetric gauged WZNW model
(obtained by  truncation of the (1,1)
supersymmetric one)  is anomalous \GRT,
i.e. is not well defined at the quantum level.

In addition to the bosonic gauged WZNW model there exist  also two other
similar non-trivial  models with corresponding \sms containing  non-zero
quantum correction terms
(but with configuration space $G$ instead of  $G/H$ one).
They are based on the  `chiral gauged' WZNW model \tye, i.e.
  are bosonic \kumar\sfet\ and
heterotic  ((1,0) supersymmetric)  \ST\  chiral gauged  WZNW  theories.
The classical action of the chiral gauged WZNW model is obtained by dropping
out the $A\A$ term in the action of the gauged WZNW model \gwznw. It reads
\gin\tye
\eqn\chi{ I_{ch} = I(g)  +{1\over \pi }
 \int d^2 z \Tr \bigl(  A\,\bd g g\inv -
 \bar A \,g\inv\del g + g\inv A g \bar A \bigr) \ .}
This action is no longer invariant under the true gauge transformations \inva\
 but is still invariant under  the    `chiral' gauge transformations
$ g \ra u\inv(z) g{ \bar u}(\bar z) $   with holomorphic and antiholomorphic
parameters. Since the `chiral' gauge
transformations do not actually eliminate  dynamical degrees of freedom
(it is  more appropriate to  consider  them as an infinite set of  global
transformations as in the standard WZNW model case \witt)
the configuration space of the \sm one  obtains  upon  elimination of the
vector gauge field from the action
is the group space $G$ and not $G/H$. Though the gauged and chiral gauged
WZNW models are closely
related  before one integrates out the vector fields,  the associated  \sms
are,
 in general, different  \sfet\ (with the exception of the
case when $H$ is abelian,  when the chiral gauged WZNW model is actually
equivalent  to a specific class  of axially gauged  $G\times H/H$ WZNW models
\ST).

Like the classical action, the effective action of the chiral gauged  WZNW
model  can be   obtained
from the effective action of the gauged WZNW model \eff\ by omitting the local
counterterm $A\A$ \sfet.
This is easy to understand by noting that in the parametrization \para\  the
action \chi\
reduces to a combination of WZNW actions  (cf. \gwzwh)
\eqn\chihc{ I_{ch} (g,A)= I(h\inv g \bh ) - I(h\inv)  - I(\bh) \ .}
The corresponding quantum effective action  is then \sfet\ (cf. \eff)
\eqn\chig{\eqalign{ \G_{ch} (g,A) &= (\kg) I(h\inv g \bh) -  (\kh )
\bigl[ I(h\inv )  +  I(\bh) \bigr] \cr
& = (\kg) \bigl( I_{ch}  (g, A) \
- \ b  \ [ \o(A)  + \W(\A)] \bigr) \ ,  \cr }    }
where  $   b \equiv - { \gG - \gH \over \kg }$  and
 $\o(A)$ and $\W(\A)$  are the same non-local functionals of $A$, $\A$
defined in  \fun.

The classical action of the (1,0) supersymmetric chiral gauged model is
obtained by replacing the fields in \chi\ or \chihc\ by  (1,0) superfields \ST.
The bosonic part of the action is
still given by \chi\ but the fermionic  contribution changes  the structure of
the effective action:
as in the (1,1) supersymmetric case \aat\  there is no quantum shift of the
level $k$ and  also there is no  $\A$-dependent term in the quantum  part of
the effective action  \ST. The
bosonic part of the resulting effective action is thus given by \ST\
(cf. \chig)
\eqn\chih{ \G_{ch}^{(1,0)}  (g,A)  =
 k  \bigl[ I_{ch}  (g, A) \  - \ b   \  \o(A)  \bigr] \ , } where
 $ b   \equiv   - {1\ov k}   {(\gG-\gH)}    .$
To find the corresponding \sms  one needs to eliminate $A,\A$ from
\chig, \chih.
It is instructive to do this by starting with the most general ansatz for the
effective action which
formally includes all the cases of the gauged and chiral gauged WZNW models
mentioned above (cf. \eff, \chig, \chih):
\eqn\gene{\eqalign { {\hat \G} (g,A) = \ \k \  \bigl\{ & I(g)
+{1\over \pi } \int d^2 z \Tr [  A\,\bd g g\inv -
 \bar A \,g\inv\del g + g\inv A g \bar A   + (a-1) A\A \ ] \cr
& - \   b\  \o (A)\  - \  \bab \ \W (\A)  \bigr\} \ , } }
where the values of the constants $\k, a,b,\bab$   are
\eqn\noa{\eqalign{
& {\rm  gauged\  WZNW}: \ \  \k = \kg \  , \ \ \ a=-b\ , \ \ \
b= \bab =   - {1\ov  \k}   {(\gG - \gH)}\  , \cr
&{\rm (1,1)\  susy\  gauged\  WZNW}: \ \ \k= k \ , \ \ \ \ a=b=\bab = 0 \ ,\cr
& {\rm chiral \ gauged \ WZNW}: \ \  \k = \kg \  , \ \ \ a=1\ ,
 \ \ \   b=\bab=  - {1\ov  \k} {(\gG - \gH)}\  ,\cr
&{\rm (1,1)\  susy \ chiral\  gauged\  WZNW}: \ \  \k = k  \  , \ \ \
a=1\ , \ \ \   b= \bab = 0 \  ,  \cr
&{\rm  (1,0)\  susy\  chiral\  gauged\  WZNW}: \  \  \k = k \  ,
 \ \ \ \  a=1\ ,   \ \ \ b =   - {1\ov  \k}   {(\gG - \gH)}\ , \ \ \  \bab =0
\ . } }
In  this section we shall use the same notation as in \notation,
\defs\ and \nvv\ with the exception that the matrices $N$, $V$ and $\V$
are now defined as
\eqn\defss{ N_{ab} \equiv M_{ab}  + a \eta_{ab} = C_{ab} +  (a-1) \eta_{ab} \ ,
\qq
 V\equiv N N^T - b\bab\ I \ , \qq  \V\equiv N^T N - b\bab\ I \ . }
Truncating the  functionals  $\o (A)$, $ \W (\A)$ in \gene\
to quadratic terms  as in \gtr\ we obtain (cf. \trun)
\eqn\truncc{\eqalign{& {{\hat \G}}_{tr}(g, A) = \k \bigl[ I (g)
  +  \D I (g, A)\bigr] \ ,\cr
&\D I(g,A) = {1\over \pi } \int d^2 z \  \big(   - A\J  -
 \bar A J   +    A  N \A
\  + \  \ha  b  \  A\tA    +   \ha \bab \  \A \tbA  \ \big)  \ , \cr } }
where $\tA$ and $\tbA$ were defined in \tatba.
Eliminating $A,\A$  we get again the non-local expression
similar to  (2.14$'$),
$$   \D I (g) =-{1\over 2 \pi }
 \int d^2 z \ \big[ J (N^TQ\inv N - b\bab Q\inv)\inv (N^T Q\inv\J - b  J) $$ $$
+ \J (NQ N^T -b\bab Q)\inv (N QJ -\bab \J)\big]   \ .
\eq{3.8'} $$
Note that in the case of the gauged WZNW model with the values of the
parameters  given by the first line in \noa\
the action \truncc\ differs from \trun\ by a total derivative term.
Since in the   case  of an arbitrary $a$ we do not have,  in general, an  extra
symmetry
analogous to the gauge invariance, we are
lacking the  principle we used in the previous section in the process of
extracting the local
part of the effective action.
One possibility is that  the resulting  ambiguity  in the expression for the
$B_{MN}$-term should  be
fixed in each particular case.
  For example,
one may expect that there should  be no extra total derivative term in (3.8)
in the chiral gauged WZNW
model case where the form of the effective action (3.3) is  fixed by the
condition of
conformal invariance (the action is expressed in terms of WZNW actions).

On the other hand, one can try again  to interpret   the parity-even nature of
the quantum term in
(3.8)  as  implying  that the antisymmetric  tensor should retain its
semiclassical
form, i.e. to adopt the `semiclassical' prescription described at the end of
the previous section.
In what follows we shall  first consider   the `corrected' prescription.

The equations for the gauge fields $A^a$, $\A^a$ that follow from \truncc\
\eqn\eqmg{ -\J  + N\A + b\ \tA   = 0 \ ,\qq
-J + N^TA  +\bab\ \tbA  =0 \ ,}
 imply  also
\eqn\eqmc{ -\tbJ  + N\tbA + b\ A + \dots  = 0 \ , \qq
- \tJ + N^T\tA   + \bab\ \A+ \dots =0 \ . }
The  local parts of the corresponding solutions  are
\eqn\solg{\eqalign{ & A_* = V\inv(N J  - \bab\tbJ) \ , \qq
\A_*= \V\inv ( N^T \J - b  \tilde J)  \ , \cr
&\tA_* = V\inv(N \tJ  - \bab \J) \ ,\qq
 \tbA_*= \V\inv ( N^T \tbJ - b   J)\ .\cr }      }
If we substitute these solutions into the action \truncc\ and  again
ignore the non-local terms we obtain the  following expression for the local
part of the
effective action \eqn\act{\eqalign{  {{\hat \G}}_{loc}(g) &= \k \bigl\{ I (g) +
 {1\over  2\pi }
 \int d^2 z \ \big( - A_*\J -  \A_* J \big) \bigr\} \cr
& = \k\ \bigl\{ I (g) +  {1\over  2\pi }
\int d^2 z \ \bigl( -2 J \V\inv N^T \J +\bab\ \tbJ V\inv \J + b\ J \V\inv \tJ
\bigr)  \bigr\}\ .\cr    }   }
This is the expression one finds by dropping out the non-local terms
in the expression (3.8')  in the most straightforward way.

The  resulting   \sm
\eqn\sx{{\hat S}(x)= {{\hat \G}}_{loc}(g)
= - {1 \over \pi \a' } \int d^2 z \ {\cal G}_{MN} (x) \del x^M \bd x^N \ ,\qq
\a' = {2\ov \k } \ , }
has a metric which is very similar to the one in \metric,
\eqn\metgen{ G_{MN} \equiv  {\cal G}_{(MN )} = { G}_{0MN }
+ 2 (  \V\inv N^T )_{ab} E^a_{(M} \E^b_{N)}
-  b  (\V\inv)_{ab} E^a_M E^b_N -  \bab (V\inv)_{ab} \E^a_M \E^b_N \ ,}
while  the antisymmetric tensor  is given by (cf. \antis, \ants)
\eqn\antgen{B_{MN} \equiv  {\cal G}_{[MN ]} = B_{0MN}
+ 2 (\V\inv N^T )_{ab} E^a_{[M} \E^b_{N]}
\  .}
The dilaton is given by \dila, but now with the definition \defss\ for the
matrix $V$.
  In general,   there is no residual gauge invariance  so that this model
 has  the group space $G$ as a configuration space.
In the case of the gauged WZNW model the  above expressions
were  given  in  \bs\aat.
Notice that the  corresponding expression for the antisymmetric tensor
\antgen\ is different  from the result \antis, \ants\ we have
found in the previous section (in particular, the   quantum
term  in \antgen\ is $O(1/k)$, not $O(1/k^2)$  as in \ants). In Section 2
 we  have   extracted the local
part of the effective action in a way that preserved gauge invariance (which is
present in the original non-local effective action \gene\ in the special case
of the gauged WZNW model).
 This gauge invariance is missing in the local part of the effective action
\act, \sx\ as we
obtained  it above, but {\it is} present in \local, \sloc.

The expressions for the \sm couplings in the case of the bosonic
 chiral gauged WZNW model (third line in \noa)
 were already found  in \sfet\ (they are given by \metgen,
\antgen\ and \dila\ with $N=C$ and $\bab=b$, see \noa).
In Section 4 we shall check (in the 2-loop approximation)
the conformal invariance of the
 \sm corresponding to the chiral gauged $SL(2,\IR)/\IR$ WZNW model.

The heterotic case (fifth line in
\noa) was treated in \ST\ with the result  \eqn\smhet{\eqalign{ &G_{MN}  =
G^{(s)}_{MN} - b
[(C^TC)^{-1}]_{ab} E^a_M E^b_N  \  , \qq G^{(s)}_{MN}={ G}_{0MN }  + 2   C^{-1
}_{ab} E^a_{(M}
\E^b_{N)}\ ,\cr & B_{MN}  = B^{(s)}_{MN}= B_{0MN} + 2 C^{-1 }_{ab} E^a_{[M}
\E^b_{N]}\  ,\cr
&\p=\p^{(s)}=\p_0 -\ha {\ \rm ln \ det \ }  C   \ .\cr  }  }
It is quite remarkable that in this case  the antisymmetric tensor and the
dilaton  retain their  semiclassical values while
the metric receives  just one  $1/k$ correction.

For all the models defined in \noa\ one can prove
 the  following theorem (see Appendix A)\foot
{This relation was  conjectured in \kir\ for the abelian $SL(2,\IR)/\IR$ coset
case
and in \BSthree\ where it was formulated in a general form for any gauged WZW
model. Subsequently,  its validity was explicitly checked for many abelian and
non-abelian cases  \BShet\bsfet\sft\BSslsu\ and proved in general for any
gauged WZNW model  \aat\ and for any chiral gauged WZNW model  \sfet.}
\eqn\theorem{ \e{-2 \phi} \sqrt{\det G_{MN}}\
= K \sqrt{\det G_{0MN}}= K({\rm Haar\ measure \ factor \ for}\ G)\ ,}
where $K$ is a constant   $k$-dependent factor given in (A.4)  (in the
 degenerate case of the  gauged
WZNW  model  one should fix a gauge and  include the Faddeev-Popov ghost
determinant  making the right hand side
of \theorem\ equal to the invariant measure factor on $G/H$). In view of the
relation of the
expression for the dilaton \dila\ to the   determinant of the matrix in the
$O(A^2)$ term  in
the effective action one can also interpret \theorem\ in the following
way: the quadratically
divergent part  of the determinant \busch\shts\
resulting from integration over  the gauge fields
combines with the
 Haar measure for the
group $G$  to give  the correct measure $\sqrt{\det G_{MN}}$  for the \sm \sx\
\tsw\aat\
  (the finite part of this determinant produces the dilaton term \wit).

In the case of the abelian  subgroup $H$  we have shown \ST\ that
the $G/H$ chiral gauged WZNW model
is equivalent to the  axially gauged $(G\times H)/H$  WZNW model
(with a special
embedding  of the subgroup $H$ into  the group $G\times H$).
Given that chiral gauged WZNW model
must  obviously be conformally invariant,  one should  fix the  ambiguity
in the  procedure of extracting of a local part of the effective action of the
corresponding axially
gauged $(G\times H)/H$  WZNW model  in such a  way that to make it
 identical to that in the case of  the chiral
gauged model.
This will be discussed on the example of the $SL(2,\IR)\times \IR/\IR$ model
in Appendix B. The conformal invariance of the resulting  \sm will be
demonstrated in  Section 4.

If one uses the `semiclassical' prescription for extracting the \sm  coupling
from (3.5)
one finds the same exact metric (3.14) but the semiclassical  antisymmetric
tensor.
For example, for the chiral gauged WZNW model the corresponding  semiclassical
$B_{MN}$  is  given by
(3.16).  This  result is in agreement with the above mentioned relation between
chiral gauged and gauged WZNW models (assuming that the same `semiclassical'
prescription is used
for both  classes of models).


\newsec{$D=3$  \sm corresponding to $[SL(2,\IR)\times \IR]/ \IR$ gauged WZNW
model:  conformal invariance at two loops}

The aim of this section is to provide a non-trivial  check
that the exact \sms  of the type  discussed  in the previous sections  are
actually
conformally invariant
beyond the semiclassical (one-loop) approximation, i.e. that the   exact
expressions for the background fields  $G_{\m\n}$, $B_{\m\n}$, $\p$
solve the string effective equations  beyond
the leading $\a'$-approximation. Such a check was already done \ts\Jack\
for the simplest exact $D=2$
\sm   corresponding to the gauged $SL(2,\IR)/ \IR$  WZNW model \dvv.
However, this $D=2$
model  has trivial antisymmetric tensor background.
In view of the subtleties associated with the
derivation of the exact expression  for $B_{\m\n}$  (see Appendix B) it is
important to  confirm
that  the exact
 backgrounds  with $B_{\m\n} \not=0$  do actually solve
the   \sm  conformal
invariance conditions.

\subsec{Description of the model}

The simplest model with  a non-trivial antisymmetric tensor coupling
is the $D=3$ `charged black string' \sm associated with  the
$[SL(2,\IR)\times \IR]/ \IR$ gauged WZNW model \horne.
The exact expression for its metric, dilaton \sft\ and  the non-vanishing
component of the antisymmetric tensor (B.17) can be represented in the form
$$ ds^2 = -{z-q-1\ov z+b } dt^2 + {z-q\ov z } dx^2  + {dz^2 \ov 4(z-q-1)(z-q)}
\ ,
 \eq{4.1} $$
$$\p=\p_0 - {1\ov 4} \ln [z(z+b)] \ , \eq{4.2} $$
$$ B_{tx} =  -
{[q(q+1+b)]^{1/2}\ov 1+b}  \big[{q+1\ov z} - {q+b\ov z +b}\big] \  \ , \eq{4.3}
$$
where
$$  q\equiv  q_0(1+b) \ ,  \ \ \ \a'= {1\ov \k}\ , \ \ \  \k=k-2\ , \  \ \
b=  - {1\ov 2 \k}
{(c_G-c_H)}= {2\ov \k}=  2\a'\ ,
\eq{4.4} $$
and $q_0$ is a free parameter  related to the  coefficient    which  determines
 the
embedding  of the subgroup $\IR$ into $SL(2,\IR)\times \IR$.
 We   discuss  the derivation  of (4.1)--(4.3) in Appendix B.  Eq.(4.3) is
found
if one uses the `corrected' prescription of Section 2 and fixes an ambiguity in
the  expression for
$B_{tx}$  in a particular  way.
In case one adopts the `semiclassical' prescription (see (2.41),(2.43)) instead
of (4.3) one finds
the following expression   (see (B.17$'$))
$$ B_{tx} =  -
{[q(q+1+b)]^{1/2}\ov (1+b)[(1+b) z- bq] }  \ . \eq{4.3'} $$
In (B.15)  we  explain the relation of the coordinates $z,x,t$ to the
`classical' group space  coordinates.\foot{In  the notation of this Section
the metric
(and the antisymmetric tensor) or $\a'$ are rescaled by a factor
of 2 as compared to the previous Sections 2 and 3 (cf.(2.23)). Note also that
the relation of our
present notation to  the notation  used in \sft\bs\sfet\ is:   $b =\lambda=
2/k', \ z= (\l + 1)  r -
\l \, \ q_0=\r^2 , \ k' = \kappa$. }  Because of the coordinates (and the
definition of  $q$) used
the `semiclassical' expression (4.3$'$) still contains  $O(b)$ corrections.
It is easy to see that the two alternative expressions (4.3) and (4.3$'$)
coincide
to the order $O(b^2)$ and thus cannot be distinguished at the two-loop order,
i.e. if the background
(4.1)--(4.3) is conformal invariant, the background (4.1),(4.2),(4.3$'$) is
conformal invariant as
well.

As was shown in \ST, the  axially gauged   $[SL(2,\IR)\times \IR]/\IR$  WZNW
model with the embedding
parameter $q_0=-\ha$ is equivalent to the   $SL(2,\IR)/\IR$ chiral gauged WZNW
model.
According to \metgen, \dila\ the \sm corresponding to the latter theory  has
the same metric and
dilaton as in  (4.1), (4.2)  \sfet\ with $q$ now fixed to be
$$  q_0= q^{(ch)}_0=-\ha \ , \ \ \ \  q= q^{(ch)} =-\ha (1+b) = -\ha {k\ov k-2}
\ , \eq{4.5} $$
i.e.
$$ (ds^2)^{(ch)} = -{[z-(1-b)/2 ]\ov z+b } dt^2 + {[z+(1+b)/2]\ov z } dx^2  +
{dz^2 \ov
4[z-(1-b)/2][z+(1+b)/2]} \ .
 \eq{4.6} $$
The antisymmetric tensor (given by (3.15)) \sfet\foot{ The imaginary unit $i$
in (4.7) can be absorbed into a
rescaling of $x$ or $t$  which will change the corresponding sign in the
metric, i.e. will change
the signature from $(- + +)$ to $(- - +)$ or $(+ + +)$ . }
  $$ B_{tx}^{(ch)} = {i{(1-b)} \ov 4} \big({1\ov z} + {1 \ov z+b } \big)     \
, \eq{4.7}  $$
is equal to (4.3)
 with  $q=q^{(ch)}$,
$$ B_{tx}(q=q^{(ch)}) =  B_{tx}^{(ch)}
     \  . \   \eq{4.8} $$
In addition to the  special case of $q=q^{(ch)}$,  the $[SL(2,\IR)\times \IR]/
\IR$ model has
two  other  obvious limits:  $SL(2,\IR)$  group space  and the direct product
$[SL(2,\IR)/\IR ]\times
\IR$ (or `neutral black string'). The two limiting \sms are already known to be
conformally invariant    so it is  useful to keep them in mind  while
analyzing  conformal
invariance of the  general  background  (4.1)--(4.3).
 The first  limit corresponds to  $q_0 =q= \infty $ and  the second -- to
$q_0=q=0$
(see also Appendix B).
Introducing the new coordinates $t',x',z'$  according to
$$ t = \sqrt q t'\  , \ \ \ \ x = \sqrt q x'\  , \ \  \ \ z = z' + q + 1  \ ,
\eq{4.9} $$
and  taking the limit $q\ra \infty$  ($z/q \ra 1 $) one finds  $$ ds^2 = -{z' }
d{t'}^2 + {(z'+1 ) } d{x'}^2  + {d{z'}^2 \ov 4z'(z'+1)} \ ,
 \eq{4.10} $$
$$ B_{t'x'} = {  z' } \ , \ \ \  \ H_{t'x'z'}=1 \ , \ \ \ \p= \const \ ,
\eq{4.11} $$
where in $\Bmn$ we have dropped out  an infinite constant (this would be
unnecessary if we
have  kept a constant term in the  derivation of (B.9) from (B.6)).
 This background corresponds to the
$SL(2,\IR)$ WZNW model. An equivalent model  is found using  the following
generalization of the
transformation (4.9)  (the case of $n= 1/2 $  was discussed in \sft):  $ t =
q^{(1-n)/2} t'\  ,  \ x
= q^{(1-n)/2} x'\  ,  \ z =q^n z' + q + 1  \ .  $ Then  assuming that  $ 0<n<1$
 and taking  the
limit $q\ra \infty $ we get instead of (4.10) $\  ds^2 = -{z' }( d{t'}^2 -
d{x'}^2)  +
{d{z'}^2/4{z'}^{2}} .$
The $q\ra 0$  ( $q_0\ra 0$) limit of (4.1)--(4.3) has the form
$$ ds^2 = -{z-1\ov z+b } dt^2 +
dx^2  + {dz^2 \ov 4z(z-1)} \ , \ \ \ \p=\p_0 - {1\ov 4} \ln [z(z+b)] \ , \ \ \
 B_{\m\n} = 0 \ ,  \eq{4.12} $$
that  represents  the direct product of the exact `black hole' background \dvv\
and
the $x$-line.
 Let us note also that  the  large $z$ asymptotics of the background
(4.1)--(4.3)  (or  of (4.12))
is described by the flat space and linear dilaton
 $$ ds^2 = - dt^2 +
dx^2  + {dy^2 } \ , \ \ \ \p=\p_0 - y \ , \ \ \
 B_{\m\n} = 0 \ ,  \ \ \ z= \e{2y} \ . \eq{4.13} $$

\subsec{Two-loop  string  effective action  and field redefinitions  }
The general strategy of a   proof  that  a background $\vp^i= (G_{\m\n},
B_{\m\n}, \p)$  solves
(in the two-loop or $\a'^2$ approximation) the \sm
conformal invariance conditions ${\bar \b}^i=0$  is the following.
One should expand the background fields in $\a'$
$$ \vp^i= \vp^i_1 + \a' \vp^i_2 + \dots \ \  ,  \eq{4.14}     $$
 and first check that the  equations
$${\bar \b}^i= {\bar \b}^i_1 (\vp)  + \a' {\bar \b}^i_2(\vp)  +\dots =0  $$
  are satisfied to the leading order in
$\a'$, i.e. in the one-loop approximation, $ {\bar \b}^i_1 (\vp_1) =0 $.
  Then one needs to find the
most general form of the two-loop ${\bar \b}^i$, e.g. by starting with the
${\bar \b}^i$-functions
computed in a particular scheme and making  the   most general local
redefinition of couplings. The
resulting ${\bar \b}^i$  will be parametrized by a number of free parameters
$d_r$.
The problem is then to show that there exists such a  `scheme', i.e. a choice
of the   parameters
$d_r$, that the equations ${\bar \b}^i=0$  are satisfied at the next order,
i.e.
$$ ({\del{\bar \b}^i_1 \ov \del \vp^j}(\vp_1) )\vp^j_2 +  {\bar \b}^i_2({\vp_1}
) =0 \ .  \eq{4.15}
$$
If $\Phi^{ir}(\vp)$ is a basis in the set of all possible
 covariant terms  with the tensor structure of $\vp^i$  constructed out of
two derivatives  of $\vp^i$  then the general
coupling redefinition $$ {\vp^i}' \ra  \vp^i + \a' d_r\Phi^{ir} +\dots  $$
induces the following change  in  the two-loop term in ${\bar \b}^i$
$$ {\bar \b}^i_2 \ra {\bar \b}^i_2 + ({\del{\bar \b}^i_1 \ov \del \vp^j})  d_r
\Phi^{jr} \ ,  $$
i.e. eq.(4.15) changes to
$$ [{\del{\bar \b}^i_1 \ov \del \vp^j}({\vp_1})] [ \vp^j_2 +  d_r
\Phi^{jr}({\vp_1})]  +  {\bar
\b}^i_2 ({\vp_1}
) =0 \ .  \eq{4.16}
$$
Therefore,  we can ignore such field redefinitions that  vanish on the
solution of the leading
order  equations.

 Equivalent approach is
to  start with the effective action
 that generates ${\bar
\b}^i$-functions,    make the  general
field redefinition   there and then take the variational derivatives.\foot {If
the effective action
is known (e.g.  is  determined from the string $S$-matrix)   then one  can
by-pass the problem of
computing independently the `diffeomorphism' vector $W_\m$ that appears  in the
Weyl anomaly
coefficients \ttt. Let us use this occasion to  correct a confusion in \met\
concerning the two-loop
value of $W_\m$.  Since the standard $\b$-function is computed only modulo the
diffeomorphism terms,
 eq.(5.35) of \met\
should depend on the scheme parameters $p_1, p_2, f_1$ only  through  the two
combinations:$\
p_1'=p_1 - \fourth f_1 ,\  p_2$ (the derivative $p_1$-term in (5.35) should
have coefficient
$p_1- \fourth f_1$). Then the
correct expression for $W_\m$  in the scheme $p_1'=-\fourth, \ p_2=0$  (see the
discussion after eq.
(6.7) in \met)   should be $W_\m= \fourth   D^\l (H^2_{\l\m}) - {5\ov 24} D_\m
H^2 $. }
  There
exists a simple scheme in which the order $\a'^2$ effective action has the form
\met\foot{More
general  actions related to (4.17) by field redefinitions are discussed in
\hult\jaj\osb.}
$$  S =  \int d^{D}x \sqrt
{\mathstrut G} \ {\rm e}^{- 2 \phi} {\tilde \b}^\p   = \int d^{D}x \sqrt
{\mathstrut G} \ {\rm e}^{- 2
\phi} \lbrace {1\ov 6}  (D-C)  -\fourth  \alpha^{\prime} [R    +  4 D^2 \p - 4
(\partial_{\mu} \phi)^{2}
- {1 \over 12} { H}^{2}_{\lambda \mu \nu}]
$$ $$  - {1 \over 16} \a'^2 [  R_{\m\n\l\k}^2 -\ha R^{\m\n\k\l}H_{\ \m\n}^\r
H_{\k\l\r}
+ {1\ov 24} H_{\m\n\l}H^\n_{\ \r\a}H^{\r\s\l}H_\s^{\ \m\a}  - {1\ov 8}
(H_{\m\a\b}H_\n^{\ \a\b})^2] +
O(\a'^3)   \rbrace \ , \eq{4.17} $$
where $H_{\m\n\l} = 3 \del_{[ \m } B_{\n\l ]}$ and $C$ is the total central
charge (we shall
ignore the trivial  part  $C_0=26$ of  $C$). The corresponding leading-order
equations  are
linear combinations of  $$   {\bar \b}^G_{\m\n} = R_{\m\n} - \fourth
H_{\m\a\b}H_\n^{\ \a\b} + 2D_\m
D_\n \p  = 0 \ , \eq{4.18} $$ $$ {\bar \b}^B_{\m\n}  = - \ha D_\l H^\l_{\ \m\n}
 + H^\l_{\
\m\n}\del_\l \p =0 \ ,  \eq{4.19} $$ $$ {\bar \b}^\p =   {1\ov 6} (D-C)   + \a'
[ - \ha  D^2 \p +
(\del \p)^2 -  {1\ov 24} { H}^{2}_{\lambda \mu \nu}] =0 \ .   \eq{4.20} $$
They are  corrected by $\a'$-terms which  contain a number of free parameters
corresponding to the most general local field redefinition \met\foot{We ignore
the term $D_\m D_\n \p$
since it can be eliminated by   a coordinate transformation,
$\ \delta x^\m = d_0\a' D^\m \p\
  , \ \  \delta G_{\m\n} = 2d_0 \a' D_\m D_\n \p$ (the equations we shall study
are covariant so
that one combination of the parameters will not be present in the variation). }
 $$ G'_{\m\n} = G_{\m\n} +  \a' S_{\m\n} \ , \ \ \  S_{\m\n} \equiv  T_{\m\n}
+  G_{\m\n} X  \ , \
\ \ \eq{4.21} $$  $$T_{\m\n} =d_1 R_{\m\n}  + d_2 \del_\m \p \del_\n \p + d_3
H_{\m\a\b}H_\n^{\ \a\b}
\ , \ \  X= d_4 R + d_5 { H}^{2}_{\lambda \mu \nu} + d_6 D^2\p + d_7 (\del
\p)^2 \ , \eq{4.22} $$
$$ B'_{\m\n}  =B_{\m\n} +  \a' K_{\m\n} \ , \ \ \ \ K_{\m\n}=  d_8 D_\l H^\l_{\
\m\n}  + d_9
\del_\l \p  H^\l_{\ \m\n} \ , \eq{4.23} $$ $$
\p'=\p + \a' Y \ , \ \ \
 Y= d_{10} R + d_{11} { H}^{2}_{\lambda \mu \nu} + d_{12} D^2\p + d_{13} (\del
\p)^2 \ . \eq{4.24} $$
In view of the observation that we  can  use the leading-order equations
(4.18)--(4.20)
to simplify the field redefinition  terms, the number of  free parameters
reduces to  10.
Moreover, in the  case of our interest the target space dimension is $D=3$ so
that
$$
H_{\m\n\l} = H \ep_{\m\n\l} \ , \ \ \ H^2_{\m\n}\equiv H_{\m\a\b}H_\n^{\a\b} =
2 H^2 G_{\m\n} \ , \ \
\  { H}^{2}_{\lambda \mu \nu}=6H^2\ , \ \ \eq{4.25}  $$
where  $\ep_{\m\n\l}$
 is the  totally antisymmetric tensor and the metric is assumed to have
euclidean signature. We
have also
$$  R_{\m\n\l\k}^2= 4R^2_{\m\n} - R^2 \ , \ \
R^{\m\n\k\l}H_{\m\n}^{\ \ \r} H_{\k\l\r} = 2 RH^2 \ , \ \  $$ $$
H_{\m\n\l}H^\n_{\r\a}H^{\r\s\l}H_\s^{\m\a}= 6 H^4 \ ,  \ \
(H_{\m\a\b}H_\n^{\a\b})^2 = 12 H^4 \  .   $$
As a result,  the  relevant field redefinitions are given by (4.21)--(4.23)
with
$$T_{\m\n} =c_1 R_{\m\n}  + c_2 \del_\m \p \del_\n \p \ , \ \ \
X= c_3 H^2 + c_4 R + c_5 \ , \ \ \eq{4.26} $$
$$  K_{\m\n}=  c_6 \ep_{ \m\n\l } D^\l H  \ , \ \ \ \ Y = c_7 H^2 +  c_8 R +
c_9\ . \eq{4.27}
$$  The  constant terms in $X$ and $Y$  appear as a  consequence of the
elimination of
the $(\del \p)^2$ terms using (4.20).
We can  set  $c_1=0$ (in view of (4.18)--(4.20)  and the previous footnote
$c_1$ can be absorbed into $c_3, c_6$, $c_7,c_8,c_9$) and $c_9=0$ (our
equations will depend
on $\p$ only through its derivatives).
We are thus left with 7 free parameters $c_2, \dots, c_8.$

To simplify the analysis,  we shall consider only  the following 3  scalar
equations (this turns out
to be sufficient in order to  prove  that our background is a solution):
the trace of the metric $\bar \b$-equation, the equation for the antisymmetric
tensor (in $D=3$ it is
equivalent to a scalar equation) and the equation for the dilaton.
These  equations have the following form   in the scheme  in which  the
effective action is given
by (4.17) \met\ (we  make use of $D=3$, (4.25) and (4.18)--(4.20))
$$ {\bar \b}^G = R- {3\ov 2} H^2  + 2 D^2 \p +
\ha  \a' [ R^2_{\m\n\l\k}   - 6RH^2 + {9\ov 4} H^4
+ 5 (D_\m H)^2 - \ha D^2 H^2 ]  + \a' Q =0 \ , \eq{4.28} $$
$$ {\bar \b}^B_{\n\l}\ep_{ \m}^{\ \n\l } = - \e{2\p}  \ \del_\m \{ \e{-2\p} [ H
  + \fourth \a'  (2RH
+ 5H^3)  + \a' F ] \} = 0   \ , \eq{4.29} $$  $$ {\tilde \b}^\p = {\bar \b}^\p
-\fourth
{\bar \b}^G
 ={1\ov 6} (D- C) - \fourth \a' \big[   R    +  4D^2 \p - 4
(\partial \phi)^{2} - {1 \over 2} { H}^{2} $$ $$ +
\fourth \a' ( R_{\m\n\l\k}^2 - RH^2 -{5\ov 4} H^4)  + \a'  P\big] =0\ .
\eq{4.30} $$
The terms  $Q,F $ and $P$ (that vanish   in the scheme of (4.17))
are included to indicate corrections which  will appear once a general field
redefinition
(4.21)--(4.24),(4.26),(4.27) is performed  in the leading-order terms
(we use again (4.18)--(4.20))
$$ Q=  D_\m D_\n S^{\m\n}  - D^2 S  + H^2 S  - 2 D_\m S^{\m\n}  \del_\n \p
 + \del^\m S \del_\m \p + 2 D^2 Y - 3 c_6 H D^2 H \ , \  \eq{4.31}  $$
$$ F=  - \ha SH - 2YH  + c_6 D^2 H  \ , \ \ \ \ \ \  S\equiv S^\m_\m= T + 3X \
,\eq{4.32}  $$
$$P=  D_\m D_\n S^{\m\n}    + S^{\m\n} (-2 D_\m D_\n \p + 4 \del_\m \p \del_\n
\p)    - 4
D_\m S^{\m\n}  \del_\n \p + 2 \del^\m S \del_\m\p  $$ $$
 +  4 D^2 Y  - 8 \del^\m Y \del_\m\p   - c_6 H D^2H  \   . \eq{4.33} $$


\subsec{Solution of  two-loop  conformal invariance  conditions}

Let us  now compute various terms in (4.28)--(4.33) in the special case of
(4.1)--(4.3), expanding the
leading $O(\a')$ terms to the first power in $\a'$ or $b$.  The necessary
geometrical  quantities
and their expansions in powers of $b$
 are given  in Appendix C.  Since  $q$ in (4.1)--(4.4) is a free  parameter it
is not necessary to
expand it in $b$ (i.e. we  can   treat it as being $b$-independent).
For generality we have done the computation  of (4.28)--(4.30) for arbitrary
values of the
constants  $s_1,s_2$  which are used to parametrize  $B_{tx}$ in (C.9), i.e.
$$ B_{tx} =  -
{[q(q+1)]^{1/2}\ov z }  \big[1 + bv_1 + { bv_2 \ov z}  + O(b^2) \big] \  \ ,
\ \ \ v_1= -\ha s_1 (1+q)\inv\ , \ \ \ v_2= \fourth (s_2-1) \ .
\eq{4.34}
$$
The antisymmetric tensors (4.3) and (4.3$'$)  coincide to this order  and  both
have
$$ s_1 = 3 + 4q\ , \ \ \ \ \     s_2=  1 + 4q \ .  \eq{4.34'} $$
  Introducing the notation
$$ I_1 = (2q+1)z^{-1}\  , \ \ \ I_1^2= (2q+1)^2z^{-2}\ , \ \ \  I_2 =
q(q+1)z^{-2}
\ , \ \ $$ $$ \ I_3= I_1I_2=  q(q+1)(2q+1)z^{-3}\ , \ \ \
I_4= I_2^2= q^2(q+1)^2z^{-4} \ ,  $$
and  using  (C.9)--(C.15) we get  ($b= 2\a'$)
$$R- {3\ov 2}  H^2  + 2 D^2 \p = b [- (4 + 6q +6qs_1) z^{-2} + (6+ 6s_2) q(q+1)
z^{-3} ] +
O(b^2)  \ ,  \eq{4.35} $$
$$ \ha  \a' [ R^2_{\m\n\l\k}   - 6RH^2 + {9\ov 4} H^4
+ 5 (D_\m H)^2 - \ha D^2 H^2 ]  =
\fourth b [ 16 z^{-2}  -16  I_2 + 80 I_3  -240  I_4 ] + O(b^2)   \ , \eq{4.36}
$$
$$ \e{-2\p}  H =    2i[ 1- \ha b s_1 (1+q)^{-1} + \ha b (1+s_2) z^{-1} ] +
O(b^2) \ ,  \eq{4.37} $$ $$  \fourth \a' \e{-2\p}  (2RH + 5H^3) = 2ib
(6 I_2 - I_1) + O(b^2)  \ , \eq{4.38} $$
$$   R    +  4D^2 \p - 4(\partial \phi)^{2} - {1 \over 2} { H}^{2}
= -4 + 2b  [- (q+1 + qs_1)z^{-1}  + (1+s_2) I_2] z^{-1}  + O(b^2) \ , \eq{4.39}
$$
$$\fourth \a' ( R_{\m\n\l\k}^2 - RH^2 -{5\ov 4} H^4) = 2b [ z^{-2} +
 2I_2 - 2 I_3] + O(b^2) \ .  \eq{4.40} $$
We have fixed $\p_0$ so that $\e{-2\p_0}  [q(q+1)]^{1/2} =1$.
Calculating  (using the software package Mathematica)
similar expansions in powers of $1/z$  for $P,F,Q$ in
(4.31)--(4.33) and combining them with  (4.35)--(4.40) we have found
that  the  leading-order terms in  the conformal invariance equations (4.28),
(4.29) and (4.30)  take
the following form
$$ {\bar \b}^G = 2\a' \big[ e_1 (1+2 q) z^{-1} + 2 e_2 z^{-2} +
e_3 q(q+1)z^{-3} + 2 e_4 q^2(q+1)^2 z^{-4}\big]  + O(\a'^2) \ ,
\ \ \eq{4.41} $$ $$
 {\bar \b}^B_{tx} =  2\a' G^{zz} {\sqrt G} \e{2\p} \del_z \big[ t_1 z^{-1} +
t_2 q(q+1) z^{-2}\big]
+ O(\a'^2) \ , \eq{4.42} $$  $$ {\tilde \b}^\p = {\tilde \b}^\p_0  -{1\over
2}\a'^2 \big[ h_2z^{-2} +
2 h_3 q(q+1)z^{-3} + 12 h_4 q^2(q+1)^2z^{-4}\big] + O(\a'^3) \ , \eq{4.43} $$
where
$$e_1= c_2 - 12 c_4 + 16 c_8 \ ,
\ \ \ e_2= 14 c_4 - 16 c_8 + (-5 - c_2 + 28 c_3 + 154 c_4 - 3 c_5 $$ $$
 + 12 c_6 - 32 c_7 -176 c_8 -
3 s_1) q + (-2 - c_2 + 28 c_3 + 154 c_4 - 3 c_5 + 12 c_6 - 32 c_7 -176 c_8) q^2
 \ , $$
$$
e_3= 26 - c_2   - 88 c_3   - 376 c_4  - 48 c_6 + 96 c_7 + 384 c_8 + 6 s_2   +
(40  - 2 c_2
 - 176
c_3    - 752 c_4   $$ $$  - 96 c_6  + 192 c_7  +768 c_8 )q \ ,
\ \ \  e_4= -30 + c_2   + 72 c_3
+ 252 c_4  + 36 c_6 - 64 c_7 - 224 c_8 \ , $$
 $$t_1= 6 + c_2   - 12 c_4  - 16 c_6 - 16 c_8 + 2 s_2   +
(8  +  2 c_2    - 24 c_4   -32 c_6  - 32 c_8)q \ , $$ $$
 t_2= -24 - c_2   + 12 c_3   + 42 c_4  + 24 c_6 + 16 c_7 + 56 c_8 \ , $$
$$h_2 =  -c_2   + 24 c_4  - 32 c_8 + (2 - 6 c_2   + 40 c_3   + 236 c_4  - 2 c_5
+ 8 c_6 -64 c_7 - 352
c_8 - 2 s_1  ) q  $$ $$  +(4 - 6 c_2   + 40 c_3   + 236 c_4  - 2 c_5 + 8 c_6 -
64 c_7 - 352 c_8) q^2
\ , $$ $$
 h_3 = -1 + c_2   - 40 c_3   - 164 c_4  - 8 c_6$$ $$  + 64 c_7 + 256 c_8 + s_2
  +( -4  + 2
c_2    - 80 c_3    - 328 c_4   - 16 c_6  + 128 c_7  + 512 c_8) q\ , $$ $$
\ \ \  h_4 = 10 c_3   + 35 c_4  + 2 c_6 - 16 c_7 - 56 c_8\ ,  $$
 $$ {\tilde \b}^\p_0 = {1\ov 6} (D- C)  + \a'  + \a'^2 h_0 + O(\a'^3) \ , \ \
h_0=  -  c_2 -
c_5   \ , \eq{4.44} $$ where $c_i$ are the scheme dependence parameters  and
$s_1,\ s_2$   appear in
$H$ in (C.9).

The conformal invariance conditions $$ e_1=e_2=e_3=e_4=t_1=t_2=h_2=h_3=h_4 =0$$
(the equation
$h_0=0$ that determines the value  of central charge will be discussed below)
is  a system of 9
equations, while the scheme ambiguity is  represented only by   7 parameters.
That is why the
existence of a solution is non-trivial.
 For generic  $q\not=0,-1$ a  solution   exists only  if   $s_2 = 1 + 4q$
which is the value
corresponding to (4.3)  (i.e. for the  special value (B.5) of the parameter
$\mu$  in
(B.9),(B.15),(C.9),(C.10)) and is given by  $$c_2 =c_4=c_8 = 0\ , \ \ \ c_3
=c_6  = \ha \ , \ \ \ c_7=
{3\ov 8} , \ \  \ c_5 = {1  + 2q - s_1\ov 1+ q}=-2 \ , \eq{4.45} $$
where we have finally set $s_1$ to its value   $s_1=4q+3$ in (4.3).
We conclude that there exists a `conformal' scheme in which the
background (4.1)--(4.3) is a solution of the
conformal invariance equations (4.28)--(4.29) in
the 2-loop approximation.  Equivalently, a background  which is related to
(4.1)--(4.3)  by the  field redefinitions  (4.21)--(4.24),(4.26),(4.27)  with
parameters given by
(4.45) is a  solution of the string  equations
 in the    scheme  where  the effective action  has the form (4.17). The
transformation  that
defines   the `conformal' scheme in terms of  that of (4.17) is  thus  simply
($c_5=-2$)
$$ G'_{\m\n} = G_{\m\n} + \fourth \a'  H^2_{\m\n}  +c_5 \a' G_{\m\n}  +O(\a'^2)
$$ $$ =
 G_{\m\n} + \ha \a'  H^2_{\m\n}  +  \a'D^2\p  G_{\m\n} - 2\a' (\del \p)^2
G_{\m\n}  +O(\a'^2)\ ,
\eq{4.46} $$
$$  B'_{\m\n} = B_{\m\n} + \ha \a' D^\l H_{\l\m\n} + O(\a'^2) \ , \ \ \
      \p'=\p + {1\ov 12}\a' H^2_{\m\n\l}  + {3\ov 8}\a' R
+ O(\a'^2)\ .  $$
In (4.46) we have made use of the leading-order equation (4.20) and
replaced $H^2 G_{\m\n}$ by  $\ha H^2_{\m\n}$.
For $\p=0$ eq.(4.46) is equivalent  to   the transformation in \met\  between
the
 scheme (``$f_1=1$") corresponding to (4.17)  and the  `conformal'  scheme
(``$f_1=-1$")
in which the  parallelizable space (e.g. a group space) is automatically a
solution
of the conformal invariance equations.  Note that the solutions of the
equations in the different
schemes are related by  the inverse transformations, i.e. if  $G_{\m\n}$ is the
solution
in the scheme (4.45) then $G'_{\m\n}$ in (4.46) is the solution in the  $c_i=0$
- scheme
of (4.17).

In particular, the background corresponding to the chiral gauged
$SL(2,\IR)/ \IR$  WZNW model (which is the special case  of (4.1)--(4.3)
for $q=-\ha(1+b)$) is  also conformally invariant.\foot {A careful analysis of
the system of
the conformal invariance equations (4.43)--(4.45) in the limit $q=-\ha + O(b)$
gives the following
solution (which again exist only if $s_2=1+4q=-1 + O(b)$ but is more general
than (4.45) with
$q=-\ha$): $c_2 = 0\ , \ \ c_3= \ha - {7\ov 4} c_4\ , \ \  c_6  = \ha \ , \ \
c_7={3\ov 8} - {7\ov 2}
c_8\ , \   \ c_5 = - 2s_1  . $}
  In the two other special  cases: $q=0$ (direct product of $SL(2,\IR)/ \IR$
gauged WZNW
model and $\IR$) and $q\ra \infty, z\ra \infty $ ($SL(2,\IR)$  WZNW model) we
find  the solutions
for the following values of non-vanishing $c_i$
 $$ q=0\ : \  \  \ c_2=c_4=c_8=0  \ , \ \ \
h_0=0\ ,  \eq{4.47} $$
$$  q=\infty\ : \ \  \ \ 4 c_3 + 6 c_4 - c_5 =4\ , \ \ \  h_0 = 2 \ .
\eq{4.48} $$ Eq.(4.47) thus reproduces the  known  result about the  2-loop
conformal invariance of
the exact black hole background  (which is true in the `standard' or
`conformal' scheme) \ts.
 Eq. (4.48) specifies the relation between the scheme of (4.17) and the scheme
in which the
$SL(2,\IR)$ group space is a solution of ${\bar \b}^i=0$. Using that
for this  group space (cf. (C.11),(C.13)) $R=-6, \ H^2=-4$  we can put the
corresponding
redefinition of the metric into the form (cf. (4.46))
$$ G'_{\m\n} = G_{\m\n} +  \a' (c_3 H^2  + c_4 R +c_5) G_{\m\n} +O(\a'^2)=
 G_{\m\n} + \ha \a'  H^2_{\m\n}   + O(\a'^2)\ .  \eq{4.46'} $$

\subsec{Scheme dependence and the value of central charge }

The conditions of conformal invariance of the \sm  ${\bar \b}^G_{\m\n}=0, \
{\bar \b}^B_{\m\n}=0$
imply \curci\tsa\osb\ that  ${\tilde \b}^\p= {\bar \b}^\p= {\tilde \b}^\p_0 $
is a constant  which
is  proportional to the central charge.  It is natural to expect that the
latter should be
consistent with the conformal field theory expression.  It is necessary,
however, to  note
that the  correspondence between the \sm and conformal field theory results may
hold only in a
specific `conformal' scheme.

 As follows  from (4.39),(4.44)  the leading order form of (4.30) is
satisfied if
 the value of the central charge is
$$ C= D + 6\a' + O(\a'^2)  = 3[1 + 2\a'  + O(\a'^2)] \  .  $$
This is consistent with the value of the central charge of the
$[SL(2,\IR)\times \IR]/\IR$ conformal field theory
$$ C= {3k\ov k-2} = 3 + {6\ov k} + {12\ov k^2}  + O({1\ov k^3})= 3 + {6\ov k-2}
=3 + 6\a'  , \eq{4.49} $$
where we  have used  the relation  $\a'= 1/(k-2)$ in (4.4).
This suggests  that if  $\a'= 1/(k-2)$ all higher order $O(\a'^n)$
contributions to $C$
should vanish in the `conformal' scheme  in which   the whole
expression for $C$ should be given just by the `one-loop' $O(\a')$
contribution (4.49).

According to (4.44),(4.45)  the $O(\a'^2)$-term  in (4.48) has the form
$$ C= 3 + 6\a' - 6\a'^2 {1  + 2q - s_1\ov 1+ q} +   O(\a'^3) = 3+ 6\a' + 12
\a'^2 +O(\a'^3) \  ,
\eq{4.50} $$
where we have used that the value of $s_1$ that corresponds to  our background
(4.3),(C.10) is
$s_1= 3+ 4q$.   This expression  would be in perfect agreement with (4.49) if
$\a'$ were equal
to $1/k$ as in the semiclassical approximation. The conformal invariance of the
\sm  holds of course
for an arbitrary choice  of the overall coefficient $\a'$.  However,  since the
$\a'$-dependence of
the background (4.1)--(4.3) was established by starting from the gauged WZNW
effective action
which led also  to the `shifted'  value $\a'= 1/(k-2)$ in (4.4)  one would
expect to reproduce
the  conformal field theory  value of $C$ (4.49) with such $\a'$ (as indeed was
in the
$D=2$  black hole model \dvv\ts).

This, in fact, is possible  by noting that there is an additional freedom  of
rescaling $B_{\m\n}$
(or its field strenght)  by a constant factor ($1+ O(\a')$) that can be
included in the  field
redefinition ambiguity (this is the analog of the transformation $G'_{\m\n} =
G_{\m\n} + \a'
c_5 G_{\m\n} + O(\a'^2)$ in (4.21),(4.26))
$$ B'_{\m\n} = B_{\m\n} + \a'c_0 B_{\m\n} + O(\a'^2)\ . \eq{4.51} $$
By the  combined  rescalings of $G_{\m\n}$ and $B_{\m\n}$ one can effectively
rescale $\a'$.
As it is clear from (C.9),(C.10) the transformation (4.51) shifts the value of
$s_1$ to $s_1'=s_1 -
c_0 (1+q)$ or $c_5$ in (4.45) to $c'_5=c_5 +c_0$ and thus $h_0$ in (4.44) to
$-c_2-c_5 -c_0$.
To  have the zero `2-loop' term in  $C$  we thus need $c_0=-c_5$ or  (since
$s_1=4q+3$)
 $c_0= 2$.  This is the expected result since, in general,  it is clear that to
change $\a'= 1/(k-2)$
into $\a'=1/k$ one is to rescale  $B_{\m\n}$ by the factor  $k/(k-2) = 1+b =1 +
2\a'$.
The rescaled $B_{\m\n}$ in (4.3)  takes  the simpler  form
$$ B'_{tx} =  -
{[q(q+1+b)]^{1/2}}  \big[{q+1\ov z} - {q+b\ov z +b}\big]   \ , \ \ s'_1= 1+2q\
, \ \ c'_5 =0\ .
\eq{4.52} $$
In what follows we shall use the present  $D=3$  example   to clarify
further the  crucial role
of  coupling   redefinitions  or  scheme dependence in  understanding   a
relation  between conformal
field theory and \sm  results.
 If  the background (4.1)--(4.3) solves the conformal invariance conditions,
 the corresponding  central charge or ${\tilde \b}^\p = \const $
 may be computed at any point, e.g.
at $\ln z  =\infty$.
 In this limit  our background  becomes a flat space with linear dilaton
(4.13).
Assuming that  higher order  $\a'^n$ contributions to  ${\tilde \b}^\p$ are
constructed in terms of
the curvature $R$ and $H$ which vanish in this limit,  one  concludes  that
the exact value of  $C$
should  be determined  just by the `one-loop'  dilaton term,
$$ C= 6{\bar \b}^\p=  3 + 6 \a' (\del \p)^2 = 3+ 6\a' \ .  \eq{4.53} $$
This, in fact, gives the  conformal field theory  value for $C$ (4.49) under
the identification
$\a'= 1/(k-2)$ in (4.4).
This argument is true  only in a special  scheme in which
 there are no higher loop contributions to
${\bar \b}^\p$  on the flat linear dilaton  background.
 {\it A priori } such  scheme  need not necessarily be  the one  in which
(4.1)--(4.3) is  conformally
invariant.
The  general redefinition  of the metric (4.21) will  induce, e.g., higher
order
$O((\del\p)^n)$  terms in ${\bar \b}^\p$ which will {\it not }  vanish on the
linear dilaton
background and will thus  shift the value of $C$.
For example,  the 2-loop shift $h_0$ in  $C$ (4.44)
is expressed in terms of the  coefficients $c_2,c_5$ of the dilatonic
$(\del\p)^2$ terms in
$G'_{\m\n}$.
  In fact, consider the asymptotic  large
distance form  of the redefinitions (4.21),(4.24)
 $$ G'_{\m\n} = G_{\m\n} +  \a'(c_2 \d_{\m 2}\d_{\n 2}  + c_5 G_{\m\n}  ) \ ,
\ \ \ B'_{\m\n} = B_{\m\n} \ , \ \ \ \ \p'=\p + \a' d_{13} \ , \eq{4.54} $$
where we have made use of (4.13).
 As a result of (4.54)
   we get a  shift  in  the overall
scale of  the  metric,
$  G'_{22} = 1 +\a' (  c_2+ c_5) $
so that the  one-loop  expression   (4.53) now produces a two-loop correction
to $C$
$$ C= 6{\tilde \b}^\p=  3 + 6 \a' {G'}^{\m\n} \del_\m \p \del_\n \p  = 3+ 6\a'
- 6\a'^2(c_2 +c_5)
 + O(\a'^3) \ . \eq{4.55} $$
The above redefinition amounts to a constant rescaling of  (the relevant
component of)  the metric,
or  to an effective redefinition  of $\a'$.

We conclude that the  fact  that the total contribution to $C$ comes just from
the one-loop
correction   is scheme-dependent, i.e.  is   true only in a specific scheme.
Another  illustration of this  point    can be given on the example of the WZNW
 theory
or the group space \sm.  This case  is `complementary' to the  gauged WZNW or
`coset' \sm  one
where the derivatives of $G_{\m\n}$ and $B_{\m\n} $   are decreasing with
distance while that of the
dilaton is approaching a constant:
 for the group space    the derivative of the dilaton is zero  but
$R=\const, \ H =\const$   so that  $C$ is expected to receive contributions to
all orders in
$\a'$.
In fact, identifying $\a'$ with $1/k$ and computing ${\bar \b}^\p$ in the same
scheme  in
which the  two-loop ${\bar \b}^G$ and ${\bar \b}^B$-functions  naturally
vanish,  one  correctly
reproduces \met\  the   $(1/k)^2$-term in the standard conformal field theory
 expression \kz\nem\  for $C$,
$$  C= D - \a'R + \a'^2
R^2D\inv  + O(\a'^3)= { kD\ov k + {\rm g_G}} \ , \ \ \  R= D{\rm g_G}\ , \ \ \
\a'=1/k \ . \eq{4.56}
$$
This can be also seen directly from (4.28)  (with $P=0$)  in the $SL(2,\IR)$
group manifold  case ($D=3, \ R_{\m\n} = \third RG_{\m\n} = \fourth H^2_{\m\n}=
\ha H^2 G_{\m\n}$).
At the same time,
  it is possible in principle to find  such  a  `non-standard' scheme in which
the
full contribution to $C$  comes just from the one-loop  $O(\a'H^2)$  term  in
(4.20),
$$ C= D + 6\a' [ - \ha D^2 \p + (\del \p)^2 - {1\ov 24} H^2_{\m\n\l} ]
= D - \a'R
 = D - {D {\rm g_G}\ov k +{\rm g_G} }\ . \eq{4.57} $$
with all higher-order contributions being  now `hidden'  in the $H^2$-term.
This happens if one  rescales both the metric $and$ the antisymmetric  tensor
$G'_{\m\n}=(\k/ k) G_{\m\n}, \  $
 $B'_{\m\n}= (\k/ k) B_{\m\n}$, i.e.
  effectively  replaces  $\a'=1/k $ by  $1/\k$.
If one starts with the effective action of the WZNW model \tsw\  where one  has
$\k$ as an  overall
coefficient,  one is to use such a scheme in order to reproduce the standard
expression for
$C$.\foot{Such a rescaling of $B_\mn$ would be unnecessary if  the
semiclassical expression for
$B_\mn$   (B.9$'$) or (4.3$'$),(B.17$'$)  were `truly semiclassical' being
multiplied  by $k$ and
not by $\kg $ in the \sm action, i.e.  contained an extra factor of $1+b$.}
Since the group space is a particular limit
of our $D=3$ background (4.1)--(4.3)  this explains also
why we need to rescale $B_{\m\n}$  (4.51)   in order  to reproduce the
correct expression for $C$ under the identification of $\a'$ with $1/\k$.
\foot{ A  scheme in which  $C$
receives
 only the one-loop contribution may look  natural from the  conformal field
theory point of view
where
 the renormalization of $k$ originates from normal ordering   and so
is  effectively  `one-loop'. The above observations  seem to  resolve the
puzzle
discussed  in \ajj. On one hand, the exact current algebra result for $C$ in
the (1,1) supersymmetric
WZNW model contains  just one $1/k$ correction, $C= D(k- {\rm g_G})/(\k- {\rm
g_G}) = D(1- {\rm g_G}/
k)$.  On the other hand, there is no reason to expect that  four  and higher
loop contributions  to
the \sm $\bar\beta$-functions and thus to   $ {\tilde \b}^\p$ should vanish in
general in this
model (see \ajj\ and references there).  The paradox  disappears once one notes
that   the
correspondence between the conformal field theory and \sm results should hold
only in a particular
scheme; there  exists such a scheme in which the whole contribution to   $ C=
6{\tilde \b}^\p$
comes just from the one-loop term.}
\newsec{Semiclassical   background  as an exact solution of  conformal
invariance equations}
As is well known,  there exists a `standard' scheme  \hullt\met\ in which the
`semiclassical' ($\a'$-independent) group space background of the WZNW model
remains the solution
of the conformal invariance equations at each order in $\a'$ expansion (in such
a scheme
$\a'=1/k$ and the central charge $C$ receives corrections  of all orders in
$\a'$).
Similar  statement is true  for the  $SL(2,\IR)/\IR$ black hole model \tseyt:
there exists a `non-standard' scheme in which the semiclassical background
\wit\
is an exact solution. This follows from the fact that     the exact `black
hole'
background \dvv\ (that solves the conformal invariance equations in a
`standard' scheme \ts\Jack) is
related to the leading-order  one \wit\  by a local, covariant  and
background-independent field
redefinition. Given that the $SL(2,\IR)$ group space   and the   `neutral black
string'
are the two particular limits ($q=\infty $ and $q=0$) of  our general
background
it is natural  to ask if there exists such a scheme
in which the semiclassical limit of  the  background (4.1)--(4.3)
$$ ds^2 = -{z-q_0-1\ov z } dt^2 + {z-q_0\ov z } dx^2  + {dz^2 \ov
4(z-q_0-1)(z-q_0)} \  ,
  \ \ \ \
\ p=\p_0 - {1\ov 2} \ln z \ ,\eq{5.1} $$
$$    B_{tx} =  -
{[q_0(q_0+1)]^{1/2}\ov z }\   , \eq{5.2} $$
is  an exact solution of the conformal invariance equations for arbitrary
$q_0$.  As we   shall show
below, this is indeed the case in the 2-loop approximation.\foot{A different
argument suggesting
that  there  should exist  a scheme in which  the semiclassical  background in
a  related  $D=3$
model  is  exact to all orders  was  recently  given  in \giki.}

To establish  the  form of the corresponding 2-loop conformal invariance
equations  (4.28)--(4.30)
one is to omit  the contributions  in (4.35),(4.37),(4.39) which came from the
$\a'$-dependence of
the background (4.1)--(4.3).   As a result, the  constants in (4.41)--(4.43)
change by some
$c_i$-independent numbers. Setting  (combinations of) these constants to zero
gives the following system of 9 equations
(now of course $s_1=s_2=0$)
 $$ c_2 - 12 c_4 + 16 c_8 =0 \ ,
\ \ \  2 + 14 c_4 - 16 c_8 + (-2 - c_2 + 28 c_3 + 154 c_4 - 3 c_5 $$ $$
 + 12 c_6 - 32 c_7 -176 c_8 ) q_0 + (-2 - c_2 + 28 c_3 + 154 c_4 - 3 c_5 + 12
c_6 - 32 c_7 -176 c_8)
q^2_0 =0  \ , $$ $$
 20 - c_2   - 88 c_3   - 376 c_4  - 48 c_6 + 96 c_7 + 384 c_8 =0 \ , $$  $$
   -30 + c_2   + 72 c_3
+ 252 c_4  + 36 c_6 - 64 c_7 - 224 c_8=0  \ , $$
 $$ 4 + c_2   - 12 c_4  - 16 c_6 - 16 c_8 =0  \ ,  \ \
-24 - c_2   + 12 c_3   + 42 c_4  + 24 c_6 + 16 c_7 + 56 c_8 =0\ , $$
$$ 2 -c_2   + 24 c_4  - 32 c_8 + (4 - 6 c_2   + 40 c_3   + 236 c_4  - 2 c_5 + 8
c_6 -64 c_7
 - 352
c_8) q_0  $$ $$  +(4 - 6 c_2   + 40 c_3   + 236 c_4  - 2 c_5 + 8 c_6 - 64 c_7 -
352 c_8) q^2_0=0
\ ,  $$ $$
  -2 + c_2   - 40 c_3   - 164 c_4  - 8 c_6 + 64 c_7 + 256 c_8=0  \ ,
\ \ \  10 c_3   + 35 c_4  + 2 c_6 - 16 c_7 - 56 c_8=0\ ,  $$
with the  general solution (cf.(4.45))
$$c_2 =-2 \ , \ \ \ c_3=1 \ , \ \ \ c_4=c_5=c_6= 0\ , \  \ \ c_7= {3\ov 16} ,
\ \  \ c_8= {1\ov 8}  \ . \eq{5.3} $$
We also get the correct 2-loop contribution to the central charge    ($h_0=-2$
in (4.44),
cf.(4.49)) with the choice of $\a'=1/k$ appropriate for a semiclassical
background.
The redefinition (cf. (4.46))
$$ G'_{\m\n} = G_{\m\n} - 2\a' \del_\m\p \del_\n \p + \ha \a'  H^2_{\m\n}
+O(\a'^2)\ ,
\eq{5.4} $$
$$  B'_{\m\n} = B_{\m\n}  + O(\a'^2) \ , \ \ \
      \p'=\p + {1\ov 32}\a' H^2_{\m\n\l}  + {1\ov 8}\a' R  +
O(\a'^2)\ . \eq{5.5}  $$
 that relates the `non-standard' scheme (5.3) to the  scheme of (4.17) is
similar to
the one  found in the $D=2$ black hole  case in \tseyt. In particular, (5.4)
contains
the term $-2\a' \del_\m\p \del_\n \p$
that can be related   \tseyt\   to  a   non-trivial  derivative term  in
the  determinant
\shts\ resulting  from  the integration over the gauge fields in the gauged
WZNW  model.

The existence of the  schemes (4.45) and (5.3)  in which  both the  exact
(4.1)--(4.3)
and semiclassical (5.1),(5.2)  backgrounds   are solutions of the conformal
invariance equations
implies that these  two  backgrounds are
related by a local field redefinition (which is
found explicitly by combining (4.46) with (5.4),(5.5) and also taking into
account a coordinate
transformation involved).\foot{Any background related to (4.1)--(4.3) or
(5.1),(5.2) by a local field redefinition will also represent a solution of the
2-loop conformal
invariance equations in a particular scheme.}
 It is thus very
likely that the full $\a'$-dependence of the exact background  (4.1)--(4.3) can
be  generated by a
local field redefinition  of the semiclassical background (5.1),(5.2).

 This does not, however,
 imply,  that it is the semiclassical and not the exact background  that has  a
direct physical interpretation. The reason  is that the `conformal' or
`standard' scheme  associated
with the exact background is more directly related to  the corresponding
conformal field theory.  In
particular,  the  equations for the perturbations of the background (e.g. the
tachyonic
equation) have simple forms  only in the `conformal' scheme  and become
complicated  in a  `non-standard' one, i.e. after the field redefinition that
transforms the exact
background into the semiclassical one \tseyt.
The propagation   of a first-quantized string  in a given background is
described  by the
Klein-Gordon-type equations for the  string modes, i.e. by the  equations  for
the states
$$(L_0 +
{\bar L_0} ) \psi_n = N_n \psi_n \ , \eq{5.6} $$
  of the corresponding conformal field theory. The stress
tensor or the operator $L_0$ of the conformal theory can be considered as a
functional  of the
background fields $\vp^i =(G_{\m\n}, B_{\m\n}, \p , \dots )$. Its structure  is
fixed in a given
 theory and does not depend  on particular identification of the background
fields.
Namely, the equations (5.6) (and hence  their solutions $\psi_n$)  do not
change  under redefinitions
of $\vp^i$.  However, the {\it form}   of their representation in terms of the
background
fields  $\vp^i$  does depend on a particular choice of $\vp^i$.   That is why
one should be careful
to take into account  the scheme dependence in comparing  the conformal field
theory and \sm results.
For example, (5.6)   should correspond to the linearized terms in the
corresponding $\bar
\b$-function equations. The latter are scheme dependent  and so the agreement
is possible only in a
particular scheme.

Let us illustrate the above remarks on the example of the tachyon $\bar
\b$-function equation
\fri\calg\tss\
  $$ {\bar\b}^T = - \gamma T  + W^\m \del_\m T - 2T \ , \ \eq{5.7} $$
$$ \gamma =\ha \a'   \Omega^{\m\n} D_{\m} D_{\n}   + \dots \ , \ \ \ \ \ \
W_\m = \a'\del_\m \p + M_\m (G,H)
 \ , \eq{5.8} $$
$$ \Omega^{\m\n}  =  G^{\m\n} + \a' l_1 R^{\m\n}  +  \a' l_2 H^{2\m\n}    +
O(\a'^2)\ . \eq{5.9} $$
Dots in $\g$ stand for higher derivative terms which are absent in the 2-loop
approximation we shall
consider below. $M_\m$ and $\O^{\m\n}$ do not depend on $\p$ in  a `standard'
class of schemes.
In the  dimensional regularisation/minimal subtraction
scheme $l_1=0$ \calg\tss;   $l_2=\fourth$  in the scheme of (4.17) (this value
can be
found from the linear term in the dilaton $\b$-function in \met).   In the
`standard' scheme  for the
group space which is related to the scheme of (4.17) by (4.46') one thus has
\osb\
 $l_2 = -\fourth$,
i.e. for the $SL(2,\IR)$ group space  with $\a'=1/k$
$$  \a' \O^{\m\n}  =  \a' \big[ G^{\m\n}   - \fourth \a' H^{2\m\n} +
O(\a'^2)\big]
 = {1\ov k} \big[
1+ {2\ov k}  + O({1\ov k^2}) \big]G^{\m\n} = {1\ov k-2 }G^{\m\n} \ . \eq{5.10}
$$
This relation is true (in the `standard' scheme) to all orders   for a general
  WZNW model \jacjon. In the scheme (4.45),(4.46) that corresponds to the
exact  background (4.1)--(4.3) we find $$   \O^{\m\n}  =  G^{\m\n}  -  c_5 \a'
G^{\m\n}  + O(\a'^2) \
. \eq{5.11} $$ With $\a' =1/k$ and $c_5=-2 $    we reproduce the expression
(5.10).
If $\a' = 1/\k$ and one rescales $B_{\m\n}$ as in (4.52) (i.e. $c'_5=0$) then
one finds  (as in
the  case of the central charge) that the exact result $\g ={1\ov 2\k }D^2 $ is
 obtained already
in the one-loop approximation. We conclude that the exact background
corresponds to the scheme in
which the tachyon equation has the same simple form as in the coset conformal
field theory.

At the same time, making the redefinition (5.4) which relates the scheme of
(4.17) to the
`non-standard' scheme where the semiclassical background is the solution,  we
find that in the latter
scheme
$$   \O^{\m\n}  =  G^{\m\n}   + 2 \a' D^\m \p D^\n \p   - \fourth \a' H^{2\m\n}
 + O(\a'^2) \ .
\eq{5.12} $$
 Being computed on the relevant  background  (5.1),(5.2) the resulting  tachyon
equation
(5.7) is of course the same as the one in the `standard' scheme or (5.6);
however, its form  in
terms of the semiclassical background fields is non-standard.

The lesson that can be drawn from the  above discussion (see also \tseyt) is
that in string theory  it is not sufficient just to find the expression for a
few non-vanishing
background fields  that solve the conformal invariance equations in a
particular scheme.  One is also
to specify how the  equations for the propagation of a first-quantized string
(i.e. the equations for
the marginal perturbations of the background) look like in that scheme.   The
answer to the latter
question is  simplified if it is known which conformal field theory corresponds
to a given solution
of the string effective equations, i.e. to a given conformal sigma model. In
that case a preferred or
a `standard' scheme is the one in which the  equations for perturbations take a
simple  canonical form
when  expressed in terms of background fields.

 As demonstrated in \ts\Jack\ and in this paper,
 the  background fields which correspond to the
coset conformal models in the $\a'\ra 0$ limit  and  which
solve   the  all-order conformal  invariance equations in the `standard' scheme
  are non-trivial
functions  of $\a'$. It  still remains to be explained on  general grounds why
there should exist a
scheme  in which the semiclassical background is also a solution, i.e. why the
$\a'$-dependence of
the exact background can be generated from the
semiclassical background by a local covariant field redefinition.  An obvious
indication that this
may be the case is the quadratic dependence of the classical gauged WZNW action
(2.1) on the gauge
field  suggesting that there may exist a scheme in which the naive gaussian
integral over $A,\A$
gives the exact answer.\foot{ The quantum gauged WZNW theory  (2.4)
on the full configuration space $g,h,\bar h$ is obviously conformally invariant
being a combination
of the ungauged WZNW  models. The corresponding `standard' scheme is the same
as in the WZNW model case.
When one integrates out  the gauge fields (or $h, \bar h$) remaining in this
`standard' scheme then
  to maintain the conformal invariance
 the \sm couplings become $\a'$-dependent (and the dilaton coupling is
induced). Preserving the
`semiclassical' form of the \sm  couplings  corresponds to switching to a
non-trivial
`semiclassical' scheme.}
 Such an argument  does not, however,  explain why the tachyon and similar
higher-mode
equations should have a non-canonical form in this `semiclassical' scheme.

\medskip

{\bf Acknowledgements}

\no
We  are  grateful   to Yu.N. Obukhov for his crucial help with GRG computer
algebra system.
We would like to thank I. Bars for comments.
K. Sfetsos is indebted to the University of Southern California for its
hospitality during
part of this work.
A.A.Tseytlin  acknowledges the  support of SERC.

\appendix{A}{Theorem on the measure factor \theorem}


Below  we shall prove the validity of the relation \theorem\ for arbitrary
values of
the parameters $a$, $b$ and $\bab$ in \gene.
We shall essentially follow the procedure used in
\sfet\ in  the particular case of the chiral gauged WZNW model.
Let us rewrite the exact metric \metric\ in the following way
\eqn\metvv{G_{MN}=G_{0MK} \tilde G^{K}{}_{N}\ , }
where $G_{0MK}$ was defined in \orc\ and
\eqn\mettvv{\eqalign{
\tilde G^{K}{}_{N}&=\d^{K}{}_{N} +(\tilde V\inv N^T)_{ab} (E^{aK}\E^b_{N}
+E^a_{N} \E^{bK})-b (\tilde V\inv)_{ab} E^{aK} E^b_{N}
-\bab (V\inv)_{ab} \E^{aK}\E^b_{N} \cr
&=\d^{K}{}_{N} +\tilde C_{a'b'} S^{a'K} S^{b'}_{N}\ , }  }
where $S^{a'}_{M}=(E^a_{M},\E^a_{M})$ and the $2 \dim H\ \times 2 \dim H$
dimensional symmetric matrix $(\tilde C_{a'b'})$ is defined as
\eqn\tcc{(\tilde C_{a'b'})=\pmatrix{-b \tilde V\inv & \tilde V\inv N^T\cr
V\inv N & -\bab V\inv }\ . }
In order to compute $\det G_{MN}$ we need $\det \tilde G^{K}{}_{N}$.
We have
\eqn\detg{\eqalign{\det\tilde G^{K}{}_{N}
&=\det(\d^{K}{}_{N} +\tilde C_{a'b'} S^{a'K} S^{b'}_{N})
=\det(\eta_{a'b'} +\tilde C_{a'c'} S^{c'M} S_{b'M}) \cr
\noalign{\vskip 6pt}
&=\det \pmatrix{I- \tilde V\inv (N^T C +b I) &  \tilde V\inv( N^T + b C^T)\cr
\noalign{\vskip 2pt} V\inv (N +\bab C) & I -V\inv (N C^T +\bab I) \cr} \cr
\noalign{\vskip 6pt}
&= \det\pmatrix{\tilde V\inv & 0\cr 0 & V\inv \cr}
\det\pmatrix{(a-1) N^T -b(\bab+1) I & N^T +b C^T\cr
\noalign{\vskip 2pt} N+\bab C &(a-1) N-\bab (b+1) \cr} \cr
\noalign{\vskip 6pt}
&=\D^{{\rm dim} H}\ ({\det V\inv})^2 \det\pmatrix{b\ I & N^T \cr
- N & -\bab\ I \cr } =\D^{{\rm dim} H}\ {\det V\inv }\ , } }
where $\D=(a-1)^2 - (b+1)(\bab+1)$. The Haar measure for the group $G$ is
given by $\sqrt {\det G_{0MN}}$ where $G_{0MN}$ is the Killing metric in \orc.
Therefore, by using \metvv, \detg\ and \dila\ one
establishes the validity of the theorem \theorem\ for  arbitrary
values of the parameters $a$, $b$ and
$\bab$ (including  the special ones  in \noa).

The case of the gauged WZNW model in  \noa\ is special:
 the above derivation formally breaks down   since the
metric $ G_{MN}$ \metric\  is degenerate (has $\dim H$ null Killing vectors,
see \zettr) as a
consequence of gauge invariance. This is also clear from (A.4): in this case
$\D=0$ and its power
$\dim H $ in (A.4)  is    the   dimension of the  null vector space of
$G_{MN}$.
Defining $\det G_{MN}$ by projecting out the zero modes by a gauge condition
one obtains \theorem\ with  the  measure on $G$   replaced by the
group-invariant measure
$\sqrt {\det (E^i_\m E^j_\n \eta_{ij}} )$ on $G/H$ (see for instance \aat ).

\appendix{B}{Exact   background corresponding  to the $[SL(2,\IR)\times \IR]/
\IR$  gauged
WZNW  model}

Below we shall derive   the  exact $\s$-model  couplings  for the
`charged black string' model \horne\sft.
The present case  with non-simple group $G=SL(2,\IR)\times \IR$  needs a
special treatment in
what concerns the antisymmetric tensor coupling:
the general expression  (2.25) derived for a simple $G$ does not apply.
To get a background that solves the (2-loop) conformal invariance conditions it
turns out to be
necessary to take into account an  ambiguity present in the  extraction  of the
local antisymmetric tensor part of the \sm action from  the effective action of
the gauged WZNW
model.

  We shall use  the following
parametrisation  for  the $SL(2,\IR)$ group element
$$ g= {\rm e}^{{i\over 2 } \t_L \s_2 } {\rm e}^{\ha
r\s_1} {\rm e}^{{i\over 2 } \t_R \s_2  } \ , \ \   \ \ \ \ \
  \  \t_L= \t + \tt \ , \ \ \ \t_R = \tt-\t \ . \eq{B.1 }
   $$
Taking $A,\A$ to be in the   axial subgroup  generated by $\ha \s_2$,
 the classical gauged WZNW action (2.1)  can be represented in
 the form\foot{The normalisation for the trace of the square of the generators
we use  below differs
by a factor of 2 from that used in Sections 2 and 3.  As a result,
the relation  between  $\a'$ and $\kappa= k -2$  we  get here and use in
Section 4
is $\a' = 1/\kappa$  and not  $\a' = 2/\kappa$  as in (2.23) or (3.13).}
$$ S (g,A) =kI(g,A) = {k \over 2 \pi } \int d^2 z [ \  L_0 (r, \t_L, \t_R)
  - A\J  - \A J - (C+ 1+ 2q_0) A\A  \  ] \  ,
  \eq{B.2}  $$
  $$ L_0 = \ha  (\del r \bd r  -  \del \t_L \bd \t_L
 -  \del \t_R \bd \t_R - 2 C \ \bd \t_L \del \t_R ) \ , $$
$$    C=C(r)  \equiv  \cosh r \ , \ \ \
J= \del \t_L +   C  \  \del \t_R \ , \ \ \
 \J = \bd \t_R + C \ \bd \t_L \ .  $$
We have added the  term
$$ - {k\pi } \int  d^2 z (\del y + \r A)(\bd y + \r \A) \ , \ \ \ \ q_0 = \r^2
$$
  with  an
 extra scalar degree of
freedom $y$ coupled to $A,\A$  and then gauged it away ($ \r$ is the   free
parameter of
embedding of the subgroup).
 The action in (B.2) is thus written already in   a particular gauge ($y=0$).
The effective action is found  in the similar way as  in (2.5)--(2.9) and  up
to a total
derivative is   given by
$$ \G(g,A)  = {\k \over 2 \pi } \int d^2 z \ [ \  L_0   - A\J - \A J -
(C+1+2q+b) A\A
   + \ha b  A{\bd\over \del} A  + \ha b  \A{\del\over \bd} \A \  ]  \ ,
\eq{B.3} $$
$$  1 + 2q + b =(1+b)(1+2q_0) \ , \ \ \ q\equiv  q_0(1+b) \ , \ \ \ 1+b = {k\ov
\k} =
{k\ov k-2} \ .  $$
The  coefficient $k$ of the extra scalar term is not renormalized since the
subgroup is abelian so
that
in the combined  expression $q_0$ is replaced by $q= q_0 k/\k= q_0(1+b)$.
Using the notation \tatba\  we can rewrite (B.4) in the following form
$$ \G(g,A)  = {\k \over 2 \pi } \int d^2 z \ [ \ L_0  - A\J  - \A J - (C+ 1+
2q) A\A
$$ $$  - \  \ha  b  \ (A- \tbA) (\A- \tA) \   - \ \ha \mu b \ (A\A - \tA\tbA)
   \  ]  \ , \eq{B.4} $$
where we have put the quantum $O(b)$ term in the  manifestly gauge-invariant
form
(cf. (2.9)) and also included  also  a
total derivative term with a constant coefficient $\mu$. The  gauge-invariant
form of this
latter term  (before  fixing  $y=0$ as a gauge) is
$$ - {\mu b\ov 2 q_0 }   \big[ (\del y + \r A)(\bd y + \r \A) -
(\del y + \r \tbA )(\bd y + \r \tA)\big] \ . $$
 Such  term does
not change anything at the level of the full effective action but will
influence the expression for
the antisymmetric tensor term  in  the local
 part of (B.6)  derived under a specific prescription (see Section 2) of  how
to drop out the
non-local terms in the process of  elimination of  $A,\A$.

One can think that the  role of this  total derivative  term  is   to
account  for the fact that  part of the $qA\A$ term is to be taken in the
`symmetrised' $\ha(\A\A +
\tA\tbA)$ form since it originates from an `extra' $\IR$-part of the full
gauged WZNW action.
This suggests that  the coefficient $\mu $ should be proportional to $q_0$.
In fact,  a natural    choice  seems to be
$$ \mu = -2 q_0= -2q(1+b)^{-1} \ , \eq{B.5}  $$
since in  this case  the local part of (B.4) takes the same form as the {\it
classical} action (B.2)
(with $k$ replaced by $\k$)
$$ \G(g,A)  = {\k \over 2 \pi } \int d^2 z \ [ \ L_0  - A\J  - \A J - (C+ 1+
2q_0) A\A
$$ $$  -  \ha  (1+2q_0) b \ (A- \tbA) (\A- \tA)   -  q_0 b (A\tA + \A\tbA)
   \  ]  \ . \eq{B.6} $$
This was the property of the effective action  in the case of the simple group
$G$ (see (2.14))
and one may try to preserve it in the  semi-simple case as well.
The choice (B.5) is distinguished also by the fact that for $q_0=-\ha$  the
effective action (B.6)
 of the  $[SL(2,\IR)\times \IR]/\IR$ gauged WZNW model automatically  reduces
to the effective action
(3.3),(3.5) of the  $SL(2,\IR)/\IR$ chiral gauged WZNW model in agreement with
the general statement
\ST\ about the equivalence of the two  models. This equivalence is proved at
the level of the full
non-local effective actions and thus it implies the equivalence of the
resulting local \sm actions provided that
consistent prescriptions for their  derivation   are used.  Correspondence with
the chiral
gauged WZNW model is an important  consistency check: first, this model must
definitely  be conformal
invariant  since its action can be expressed as  a combination  of the WZNW
actions (3.2),(3.3), and,
second,
 its effective action does not contain the  quantum $A\A$-term and thus the
determination  of
$B_{\m\n}$  is unambiguous.
As shown in Section 4, the \sm  background fields  that correspond to (B.4)
with $\mu$  given by (B.5)
solve the 2-loop conformal invariance equations.

Starting with (B.6)  and using the `corrected' prescription of Section 2, i.e.
solving the  equations
for $A,\A$ as in (2.15)--(2.17) (note that the total derivative $\mu$-term
does not influence the
equations of motion) and  substituting the solution back into the action  (B.4)
 we get
the following \sm  action in terms of the coordinates $x^\m=(r,\t,\tt)$
 (for generality we  keep the value of $\mu$ in (B.4) arbitrary)
$$ S(r, \t,\tt)= {\k \over  \pi } \int d^2 z [   G_{\m\n }  \del x^\m \bd x^\n
+   B_{\t \tt}    ( \del \t \bd \tt  -  \del \tt \bd \t ) ] \ , \eq{B.7}
 $$
$$  ds^2 = G_{\m\n} dx^\m dx^\n=
\fourth dr^2 +  (1 + q + b)  {C-1 \ov C+ 1+2q + 2b } d\t^2
  -q {C+1 \ov C + 1+ 2q } d\tt^2 \ , \eq{B.8} $$
 $$   B_{\t \tt}   =  - (2-\mu) {q(q+1)\ov  C+1+2q} -  \mu {(q+b)(q+1+b)\ov
C+1+2q+2b}  \ ,
 \eq{B.9}
$$
 $$ \p=\p_0 - {1\ov 4} \ln [(C+1 + 2q)(C+1+ 2q +2b)]  \ . \eq{B.10} $$
As expected, the parameter $\mu$ appears only in the expression for the
antisymmetric tensor
(dependence on $\mu$ disappears of course in the classical limit $b=0$).
 Note   that in the limit $q=0$ the model (B.7)--(B.10)  reduces to
the  direct product of the $SL(2,\IR)/\IR$ and $\IR$ as it should ($\tt$
decouples in (B.8) but this
is a gauge artifact  that  can be avoided by first rescaling $\tt$ and then
taking the limit).

Eq. (B.9)  considered formally for arbitrary values of $\mu$ has several
special  cases. When $\mu=0$
in (B.4), i.e. when the  effective action  is taken in the same  form  as in
(2.14)
 we get
the  analog of (2.25)\foot{This expression found in \bs\  using a   manifestly
gauge-inavriant
prescription  does not represent  a conformally invariant background (cf.
Section 4).
Note that since
in the present case the group $G$ is not simple,
 eq. (B.11) {\it does  contain}    quantum corrections
even though the
subgroup  here is  one-dimensional  ((B.11) is the same
$B_{\m\n}$ that follows simply from  the classical action  (B.2) {\it but} with
$q_0$ replaced by
$q$).}
$$   B_{\t \tt}   = - {2q(q+1)\ov C+ 1 + 2q } \ .  \eq{B.11} $$
Another special case is $\mu=1$.  In this case
(B.6) is exactly equal to the `naive' form of the effective action (B.3) (as in
(3.8)--(3.12)). The
resulting   expression for $B_{\m\n} $
 $$   B_{\t \tt}   =  - {q(q+1)\ov  C+1+2q } -  {(q+b)(q+1+b)\ov  C+1+ 2q+2b}
\ ,
 \eq{B.12}
$$
is the same one that one finds  (after gauge fixing) from the
analog of (3.15) in the  axial  subgroup  case.  For
 $q_0= - \ha, \ q= -\ha (1+b)$   (B.12)  gives
  $B_{\m\n}$  in   the
$SL(2,\IR)/\IR$ chiral gauged  WZNW model.

The  expression for the exact antisymmetric tensor  coupling of
the  $[SL(2,\IR)\times \IR]/\IR$ gauged WZNW model  that corresponds to   $\mu$
 in (B.5) is
 $$   B_{\t \tt}   =  - {2q(q+1+b)\ov (1+b)}\big[ {q+1 \ov  C+1+2q  }  -
  { q+b\ov  C+1+ 2q+2b} \big] \ .  \eq{B.13} $$
When  $q_0=-\ha$
(i.e. $\mu=1$) not only the exact metric (B.8) and the dilaton (B.10) but also
the  exact
antisymmetric tensor (B.13) of the  $[SL(2,\IR)\times \IR]/\IR$ gauged WZNW
model  coincide with the
background fields of the
 $SL(2,\IR)/\IR$ chiral gauged WZNW model.\foot{That this  is not the case for
(B.11), i.e. for
$\mu=0$   was  a puzzle  in  \sfet.}
The expression (B.13)
vanishes for $q=0$ and in the $SL(2,\IR)$ group space limit $q_0\ra \infty$
reduces to  the  correct
result  $B_{\t \tt}= \ha C $.  The background (B.8),(B.10),(B.13) thus
consistently
includes (as special cases for $q_0= 0, -\ha, \infty$) other known $D=3$ exact
conformally invariant
backgrounds.

If one uses the `semiclassical' prescription  for computing the antisymmetric
tensor one finds that
parity-even  quantum term in  (B.6)  does not produce a quantum correction to
the antisymmetric
tensor,
i.e. the latter is given by
 $$   B_{\t \tt}   =  -  {2q_0(q_0+1)\ov  C+1+2q_0}  =   -  {2q(q+1 + b )\ov
(1+b) [(1+b)
(C+1) +2q]} \ . \eq{B.9'}
$$

Introducing the new coordinates $z,t,x$
$$  2z= C+1+ 2q\ , \ \  \  it=(1+b+q)^{1/2} \t\ , \ \ ix=-q^{1/2}\tt \ \ ,
\eq{B.14} $$
we find that the metric (B.8) and the  dilaton (B.10)
take the form
(4.1) and (4.3) while  the antisymmetric tensor  $B_{tx}$ corresponding to
(B.9) becomes
$$ B_{tx} =  {p_1\ov z} + {p_2\ov z +b} \  , \ \eq{B.15}  $$
$$ \ p_1=  -\ha (2-\mu )
(q+1)q^{1/2}(q+1+b)^{-1/2}  \ , \ \ \ p_2= - \ha \mu
(q+b)q^{-1/2}(q+1+b)^{1/2}  \  , \eq{B.16}   $$
or   for the value of $\mu$  in (B.5)
$$ B_{tx} =  - {[q(q+1+b)]^{1/2}\ov 1+b}  \big[{q+1\ov z} - {q+b\ov z +b}\big]
\  . \ \eq{B.17}  $$
The semiclassical expression  (B.9$'$)  for the antisymmetric tensor  in the
coordinates (B.14)
is given by
$$ B_{tx} =  - {[q(q+1+b)]^{1/2}\ov (1+b)[(1+b) z- bq]}  \  . \ \eq{B.17'}  $$




\appendix{C}{Geometrical  objects for the exact `black string' $D=3$
background}
Below we present the results of computation of some  geometrical quantities for
the background
(4.1)--(4.3)  and their expansion in $b$  which were found found  using  GRG
computer algebra
system \GRG.\foot{We are grateful to Yu.N. Obukhov for  helping us with  this
calculation.} All
indices below are with respect to the local vierbein  corresponding to the
metric (4.1)
(with the flat metric being $(-1,+1,+1)$).

The exact expressions for the non-vanishing components of the  Ricci tensor and
curvature scalar are
$$
R_{00}=(1 + q + b)(qb + 4qz + bz - 2z^2)[z(b + z)^2]^{-1}  \ ,  \eq{C.1}
$$
$$
R_{11}=- q(3b + 3qb + 4z + 4qz - bz - 2z^2)[z^2(b + z)]^{-1} \ ,   \eq{C.2}
$$  $$
R_{22}=(-3qb^2 - 3q^2b^2 - 6qbz - 6q^2bz + 2qb^2z - 6qz^2 $$ $$ -
    6q^2z^2 - bz^2 - b^2z^2 + 2z^3 + 4qz^3 + 2bz^3)[z^2(b + z)^2]^{-1}   \ ,
\eq{C.3}
$$
$$ R=2(-3qb^2 - 3q^2b^2 - 7qbz - 7q^2bz + qb^2z - 7qz^2 $$ $$ -
      7q^2z^2 - bz^2 - qbz^2 - b^2z^2 + 2z^3 + 4qz^3 + 2bz^3)[z^2(b +
z)^2]^{-1}\ . \eq{C.4}
$$
The vierbein components  of the second covariant derivative of the dilaton  are
$$ D_0D_0 \p =- \ha (1 + q + b)(q - z)(b + 2z)[z(b + z)^2]^{-1}\ ,\eq{C.5}  $$
$$D_1D_1\p  =\ha  q(1 + q - z)(b + 2z)[z^2(b + z)]^{-1}\ , \eq{C.6}  $$
$$D_2D_2\p  =\ha \big\{(q - z)z(b + z)(b + 2z) +
      (1 + q - z)\bigl[ z(b + z)(b + 2z)  $$ $$  -
         4(q - z) z(b + z) + 2(q-z)(b + 2z)^2 \bigr] \big\} [z^2(b +
z)^2]^{-1}\ , \eq{C.7}
$$
so that
$$
D^2\p= -\ha (-3qb^2 - 3q^2b^2 - 8qbz - 8q^2bz + b^2z + 2qb^2z - 8qz^2 -
    8q^2z^2 + 2bz^2$$ $$  + 4qbz^2 + b^2z^2 + 4z^3 + 8qz^3 + 4bz^3)
  [z^2(b + z)^2]^{-1}\ .
$$
 Also,
$$
(D_\m\p)^2=[(-1 - q + z)(-q + z)(b + 2z)^2][4z^2(b + z)^2]^{-1} \ .
$$
Computing the field strength for  $B_{\m\n}$ in (B.15),(B.17)
we get for the scalar $H$  (we use the metric with euclidean signature)
 $$ H= G^{-1/2}  H_{txz}= i{\sqrt{z(z+b)}\ov \sqrt {q(1+q+b)}} [
(2-\mu){q(q+1)\ov
z^2 } + \mu {(1+q+b)(b+q)\ov (z+b)^2} ] \  . \eq{C.8} $$
Its expansion in powers of $b$  can be represented as (since $q_0$ is a free
parameter we can treat $q$ as being $b$-independent)
$$ H^2= - 4q(q+1)z^{-2}[1 - bs_1 (1+q)^{-1}  + bs_2 z\inv ] + O(b^2) \ ,
\eq{C.9}  $$
$$ s_1= 1 - \mu (2q+1)q^{-1}=4q+3 + O(b) \ , \ \ \ \ s_2 =  1 - 2\mu = 4q+1
+O(b)   \ . \eq{C.10} $$
The form of the expansions to the first order in $b$ is (for $\mu =-2q+ O(b)$)
$$ R=4(2q+1)z\inv - 14q(q+1)z^{-2}
 +  b [ 4z\inv  + ( -10 -18q)z^{-2} + 14q(q+1)z^{-3}]  + O(b^2)\ , \eq{C.11 }
$$ $$
D^2\p
= -2(2q+1)z\inv + 4q(q+1)z^{-2}
+ b [ - 2 z\inv  +   3 (2q+1)z^{-2} - 4q(q+1)z^{-3} ] + O(b^2)\ , \eq{C.12 } $$
$$
H^2 =  -4q(q+1) z^{-2}  +  b [  4q(4q+3)z^{-2}  -  4q(q+1)(1+4q) z^{-3}]  +
O(b^2)\ , \eq{C.13} $$
$$
(D\p)^2
= 1 - (2q+1)z\inv + q(q+1) z^{-2}
+ b [ - z\inv + (2q+1)z^{-2} - q(q+1)z^{-3}] + O(b^2)
\ . \eq{C.14 } $$
To the leading order  $ H {\rm e}^{-2\p} = \const$ and
 $$  R_{\m\n\l\k}^2= 4R^2_{\m\n} -R^2  $$ $$ = 16  z^{-2} +
 32q(q+1)z^{-2} - 48 q(q+1)(2q+1)z^{-3} + 76q^2(q+1)^2z^{-4} + O(b) \ .
\eq{C.15} $$
It is
possible to show that  for any number $p$
  (note that  $\p=\p_0 - \ha \ln z + O(b)$)
$$ D^2 z^p =  z^p [ 4p^2 - 4p(p-1)(2q+1)z\inv + 4p(p-2)q(q+1)z^{-2}] + O(b) \ ,
 \eq{C.16} $$ where
$D^2 = {1\ov \sqrt G} \del_\m(G^{\m\n} \sqrt G\del_\n)$.

\vfill\eject
\listrefs
\end